\documentclass[10pt]{article}

\usepackage{float}
\usepackage{booktabs}   
\usepackage{array}  
\usepackage{cite}
\usepackage{comment}
\usepackage{amsmath,amssymb,amsfonts,amsthm}
\usepackage{acronym}
\usepackage{makecell}  
\usepackage{multirow}
\usepackage{cases}
\usepackage{graphicx}
\usepackage{textcomp}
\usepackage[font=small]{caption}
\usepackage[font=small]{subcaption}
\usepackage{makecell}
\usepackage{threeparttable}
\usepackage{color}

\usepackage{geometry}   

\usepackage[hidelinks]{hyperref}

\usepackage{setspace}
\usepackage[mathscr]{euscript}

\newcommand{\bfg}[1]{\boldsymbol{#1}}
\newcommand{\bfb}[1]{\boldsymbol{\rm #1}}
\newcommand{\T}{^{\intercal}}

\usepackage{float}

\newcommand{\jj}{\jmath}
\newcommand{\nx}{n}
\newcommand{\ny}{m}
\newcommand{\nxy}{r}
\newcommand{\xys}{\ensuremath{\mbox{$\bfb {x}$}}}

\newcommand{\jac}[2]{\bfg{{#1}}_{\hspace{-0.2mm}#2}}
\acrodef{cis}[CIS]{Continuation of Invariant Subspaces}
\acrodef{rocof}[RoCoF]{rate of change of frequency}
\acrodef{hm}[HM]{Heun's method}
\acrodef{fem}[FEM]{forward Euler method}
\acrodef{rk4}[RK4]{4-th order Runge-Kutta}
\acrodef{tm}[TM]{trapezoidal method}
\acrodef{pf}[PF]{participation factor}
\acrodef{sssa}[SSSA]{small-signal stability analysis}
\acrodef{bem}[BEM]{backward Euler method}
\acrodef{dae}[DAE]{differential algebraic equation}
\acrodef{pll}[PLL]{phase-locked loop}
\acrodef{sm}[SM]{synchronous machine}
\acrodef{der}[DER]{distributed energy resource}

\acrodef{ode}[ODE]{ordinary differential equation}
\acrodef{tdi}[TDI]{time-domain integration}
\acrodef{agc}[AGC]{automatic generation control}
\acrodef{pss}[PSS]{power system stabilizer}
\acrodef{avr}[AVR]{automatic voltage regulator}
\acrodef{tg}[TG]{turbine governor}
\acrodef{srf-pll}[SRF-PLL]{synchronous reference frame phase-locked loop}
\acrodef{gte}[GTE]{global truncation error}
\acrodef{fr}[FR]{frequency regulation}
\acrodef{cil}[CIL]{constant impedance load}
\acrodef{svt}[SVT]{single vector tracking}
\acrodef{aiits}[AIITS]{All-Island Irish Transmission System}
\acrodef{coi}[COI]{center-of-inertia}

\definecolor{blue1}{RGB}{0,60,200}

\def\BibTeX{{\rm B\kern-.05em{\sc i\kern-.025em b}\kern-.08em
    T\kern-.1667em\lower.7ex\hbox{E}\kern-.125emX}}

\theoremstyle{definition}

\renewcommand\appendix{%
  \par
  \setcounter{section}{0}%
  \setcounter{subsection}{0}%
  \setcounter{equation}{0}%
  \renewcommand\thesection{\appendixname~\Alph{section}}%
  \renewcommand\theequation{\Alph{section}.\arabic{equation}}%
  \renewcommand\thefigure{\Alph{section}.\arabic{figure}}%
  \renewcommand\thetable{\Alph{section}.\arabic{table}}%
  \addtocontents{toc}{\string\let\string\numberline\string\tmptocnumberline}%
}

\begin{document}

\begin{center}
{\large\bf On the Eigenvalue Tracking of Large-Scale Systems}

\vskip.20in

Andreas Bouterakos$^a$, 
Joseph McKeon$^a$
Georgios Tzounas$^a$$^{*}$ 
\\[2mm]
{\footnotesize
School of Electrical and Electronic Engineering,
University College Dublin, Ireland
\\[5pt]
$^{*}$Corresponding author.
\\E-mail address: georgios.tzounas@ucd.ie

}
\end{center}

{\small
\noindent
\textbf{Abstract} \   The paper focuses on the problem of tracking eigenvalue trajectories in large-scale power system models as system parameters vary.  A continuation-based formulation is presented for tracing any single eigenvalue of interest, which supports sparse matrix representations and accommodates both explicit and semi-implicit differential-algebraic models.
Key implementation aspects, such as numerical integration, matrix updates, derivative approximations, and handling defective eigenvalues, are discussed in detail and practical recommendations are duly provided.  The tracking approach is demonstrated through a comprehensive case study on the IEEE 39-bus system, as well as on a realistic dynamic model of the Irish transmission system.
}
\\[3pt]

%\vskip.1in

\section{Introduction}
%\label{sec:intro}

\subsection{Motivation}\label{motivation}

Eigenvalue analysis is a fundamental component of power system stability assessment.  It is widely used to evaluate the system's small-signal stability margin and to characterize the structure of its dynamic modes \cite{milano2020eigenvalue}. 
Within this framework, eigenvalue tracking -- the process of monitoring how eigenvalue trajectories evolve with variation of system parameters -- is a key task.
In real-world power system models, 
eigenvalue tracking typically 
involves large, non-symmetric matrices/pencils and 
can be a computationally intensive and numerically challenging problem.

\subsection{Literature Review}

The most straightforward approach to eigenvalue tracking is to perform a full eigendecomposition of the system's state matrix or pencil at each parameter value of interest.  
Since standard eigensolvers commonly reorder eigenvalues between computations, this approach must be combined with a technique to sort eigenvalues consistently at successive parameter values.  A simple approach is to pair eigenvalues based on Euclidean distances in the complex plane; however, this often leads to erroneous results, especially when different eigenvalues come close together or cross.  More robust approaches include using 
metrics based on eigenvectors, such as
participation factors \cite{tajoli2023mode} or 
modal assurance criteria \cite{MAC_Pastor_2012}. 

Repeated eigendecomposition starts to perform poorly as the system size increases. This is because QR and QZ factorizations -- the standard dense-matrix methods for computing the full spectrum of general non-symmetric eigenvalue problems -- have computational complexity and memory requirements that scale cubically with size \cite{tzounas2020comparison}. 
In practice, for large power system models involving thousands to tens of thousands of variables, computing eigenvalue trajectories becomes prohibitively expensive.  To address this, alternative solvers designed for large sparse matrices can be employed, such as Krylov-based or contour integration methods \cite{lehoucq1996deflation,stewart2002krylov,sakurai2003projection}.  
These are subspace methods which restrict the computation to only a part of the spectrum.  This is generally not a limitation for eigenvalue tracking, as in practice only the trajectories of a single or a few modes, conventionally the most critical ones for stability, are of interest.  However, the main difficulty in this case lies in the need for a proper spectral transformation (such as Cayley or shift \& invert) that can effectively target a region of the complex plane containing the eigenvalue of interest \cite{milano2020eigenvalue}.  These transformations require careful tuning at each parameter value, and maintaining consistency across a range in a reliable and automatic manner can be impractical.

An alternative approach to eigenvalue tracking is offered by the family of continuation-based methods.  The idea of continuation is to trace the solution of a problem that depends on a varying parameter, by incrementally updating it through a numerical method using previous values.  These types of techniques have been applied with great success in power system applications; for instance, they have been widely used to trace equilibria in long-term voltage stability studies \cite{ajjarapu1992continuation,canizares1993point}.  In the context of eigenvalue tracking, existing works explore continuation-based methods to follow the trajectory of either a single eigenpair at a time~[\citenum{Wen_2006}, \citenum{li2017eigenvalue}] or of multiple eigenpairs simultaneously \cite{Cheng_2011,tdps1}. The focus in these studies is on critical synchronous machine electromechanical modes.     

Despite these advances, the use of continuation-based eigenvalue tracking techniques remains limited.  For small to medium-sized systems, repeated eigendecomposition is typically preferred due to its simplicity and reasonable efficiency.  For large systems, eigenvalue tracking is most often completely avoided.  The key reason lies in how existing formulations handle differential-algebraic models:  they rely on dense matrix computations, which do not scale well thus preventing these methods from realizing what, to the authors' view, could be their main competitive advantage, namely, the ability to offer significant computational speedups in large-scale settings.  
On the other hand, existing sparse-matrix approaches, like the one in \cite{li2017eigenvalue}, rely on Newton-based tracking and can thus be sensitive to initial guesses and to trajectory discontinuities.
%resort to using Newton's method to iteratively trace eigenvalues;
%In a different approach, \cite{li2017eigenvalue} retains the sparsity of the system. However, the authors confine their study to using Newton's method to iteratively trace eigenvalues; 
%thus their results are prone to errors when the initial guess is poor or the trajectories present discontinuities. 
An additional reason is that several important implementation aspects, such as parameter step-size selection, 
matrix updates, derivatives evaluation, 
handling defective eigenvalues~\cite{zeng2016sensitivity, dobson2001strong, wang2018analysis},
have not been sufficiently addressed.

\subsection{Contributions}

The contributions of this paper are as follows:
\begin{itemize}
\item A general continuation-based tracking formulation is developed that supports both dense and sparse matrix representations, and is compatible with both explicit and semi-implicit differential-algebraic forms of large-scale power system models;  
\item The formulation is robust under large parameter step sizes, discontinuities in eigenvalue trajectories, as well as in handling  defective eigenvalues and bifurcation points.
\item Practical and critical implementation aspects -- such as multiple eigenpair tracking, parameter step-size selection,  matrix updates, and derivative evaluation, -- are addressed in detail.
\end{itemize}

These contributions are supported by a comprehensive case study. First, continuation-based tracking is applied to the \acf{fr} mode, a dynamic mode that plays an increasingly important role in analyzing system-wide oscillatory behaviors introduced by frequency control loops.  Second, it is benchmarked against repeated eigendecomposition on a large-scale model of the all-island Irish transmission system, with emphasis on computational cost and scalability.

\subsection{Paper Organization}

The remainder of the paper is structured as follows.  Section~\ref{sec:sssa} provides preliminaries on \ac{sssa}. The formulation and implementation aspects of the eigenvalue tracking approach adopted are discussed in Section~\ref{sec:eig_track}.  Section~\ref{sec:case} presents simulation results on the tracking of \ac{fr} modes, based on the IEEE 39-bus system, as well as on a real-world size model of the Irish transmission system. Finally, conclusions are drawn and future work directions are outlined in Section~\ref{sec:conclusions}.

\section{Small-Signal Stability Analysis}\label{sec:sssa}

\subsection{Power System Model}

In short-term stability analysis, power system dynamics can be studied through a set of non-linear, \textit{semi-implicit} \acp{dae}, as follows \cite{semi:2016}:
\begin{align}
\label{eq:sidae}
  \begin{bmatrix}
    \bfg T & \bfg 0_{\nx,\ny} \\
    \bfg R & \bfg 0_{\ny,\ny}
    \end{bmatrix}
      \begin{bmatrix}
       {\bfg x}'(t) \\
       {\bfg y}'(t)
    \end{bmatrix}  &=
    \begin{bmatrix}
        \bfg f( \bfg x(t), \bfg y(t) ) \\
        \bfg g( \bfg x(t), \bfg y(t) )
    \end{bmatrix} 
\end{align}
where $\bfg x(t): [0,\infty) \rightarrow \mathbb{R}^{\nx}$ are the state variables of dynamic components connected to the power network (generators, dynamic loads, automatic controls, etc.); $\bfg y(t): [0,\infty) \rightarrow \mathbb{R}^{\ny}$ are the algebraic variables (bus voltages and power injections, auxiliary variables that define control setpoints, etc.); $\bfg f : \mathbb{R}^{\nx+\ny} \rightarrow \mathbb{R}^{\nx}$ and $\bfg g : \mathbb{R}^{\nx+\ny} \rightarrow \mathbb{R}^{\ny}$ are functions that define, respectively, the differential and algebraic equations; $\bfg T \in \mathbb{R}^{\nx \times \nx}$, $\bfg R\in \mathbb{R}^{\ny \times \nx}$;
and $\bfg 0_{\nx,\ny}$ denotes the zero matrix of dimensions $\nx \times \ny$.
Note that \eqref{eq:sidae} preserves the separation between differential and algebraic variables, consistent with \ac{dae} theory: algebraic variables may vary in time but are not governed by differential equations and, by definition, do not have nonzero finite time constants \cite{milano2022power}. 
The semi-implicit \ac{dae} form \eqref{eq:sidae} is a generalization of the conventional, \textit{explicit} \ac{dae} form \cite{kundur:94}:
\begin{equation}
\begin{aligned}
\label{eq:dae:explicit}
{\bfg x}'(t) &= \bfg f( \bfg x(t), \bfg y (t) ) \\
\bfg 0_{m,1} &= \bfg g( \bfg x(t), \bfg y (t) ) 
\end{aligned}    
\end{equation}
In the remainder of the paper we work with \eqref{eq:sidae}, which offers certain practical advantages over \eqref{eq:dae:explicit}, such as increased sparsity and convenient switching of states to algebraic variables.  The explicit form can be always retrieved as a special case, for $\bfg T = \bfg I_\nx$, $\bfg R = \bfg 0_{\ny,\nx}$, where $\bfg I_\nx$ is the $\nx \times \nx$ identity matrix.

\subsection{Eigenvalue Analysis}

We are interested in studying the system's behavior in response to the variation of certain system parameters. To this end, the parameters of interest are expressed as functions of a scalar continuation parameter $p \in \mathbb{R}$
and \eqref{eq:sidae} is rewritten as follows:
\begin{align}
\label{eq:sidae:p}
  \begin{bmatrix}
    \bfg T(p) & \bfg 0_{\nx,\ny} \\
    \bfg R(p) & \bfg 0_{\ny,\ny}
    \end{bmatrix}
      \begin{bmatrix}
       {\bfg x}' \\ {\bfg y}'
    \end{bmatrix}  &=
    \begin{bmatrix}
      \bfg f( \bfg x, \bfg y, p ) \\
      \bfg g( \bfg x, \bfg y, p )
    \end{bmatrix} 
\end{align}
where the dependency of time is omitted for simplicity.
Considering small disturbances, 
%of \eqref{eq:sidae:p}, 
and aiming at using well-established results from linear stability theory, \eqref{eq:sidae:p} can be linearized around an equilibrium $(\bfg x_o, \bfg y_o) := [\bfg x_o\T, \bfg y_o\T]\T$ ($\T$ indicating the transpose) as follows:
\begin{align}
\label{eq:sidae:lin}
  \begin{bmatrix}
    \bfg T(p) & \bfg 0_{\nx,\ny} \\
    \bfg R(p) & \bfg 0_{\ny,\ny}
    \end{bmatrix}
      \begin{bmatrix}
      \Delta {\bfg x'} \\
      \Delta {\bfg y'}
    \end{bmatrix}
    &=
    \begin{bmatrix}
     \jac{f}{x}(p) & \jac{f}{y}(p)  \\
     \jac{g}{x}(p) & \jac{g}{y}(p) 
    \end{bmatrix} 
      \begin{bmatrix}
      \Delta {\bfg x} \\
      \Delta {\bfg y}
    \end{bmatrix}
\end{align}
where $\Delta \bfg x = \bfg x - \bfg x_o$, $\Delta \bfg y = \bfg y - \bfg y_o$; $\jac{f}{x}$, $\jac{f}{y}$, $\jac{g}{x}$, $\jac{g}{y}$ are the Jacobian matrices evaluated at $(\bfg x_o, \bfg y_o)$, for example, $ \jac{f}{x} = \left. \partial \bfg{f} / \partial \bfg{x} \right|_{(\bfg x_o,\bfg y_o)} $.
System \eqref{eq:sidae:lin} is of the form: 
\begin{equation}
\label{eq:singps}
  \bfb{E} (p) 
  \, \xys' = \bfb A(p) \, \xys
\end{equation}
where $\xys=(\Delta \bfg x, \Delta \bfg y)$ and
\begin{equation}
\label{eq:matrices:sidae}
  \bfb{E}(p) \equiv 
  \begin{bmatrix}
    \bfg T(p) & \bfg 0_{\nx,\ny} \\
    \bfg R(p) & \bfg 0_{\ny,\ny} \\
  \end{bmatrix}
  , \;
  \bfb{A}(p) \equiv 
  \begin{bmatrix}
   \jac{f}{x}(p) & \jac{f}{y}(p) \\
   \jac{g}{x}(p) & \jac{g}{y}(p) \\
  \end{bmatrix} \, 
\end{equation}
Alternatively to working directly with \eqref{eq:sidae:lin}, 
algebraic variables can be eliminated from the linearized system, leading to: 
\begin{equation}
\label{eq:ode:lin}
    \Delta{\bfg{x}'} = 
    [\bfg{T}^{-1}(p) \cdot (\jac{f}{x}(p) - \jac{f}{y}(p)\cdot (\bfg{g}_y(p) - \bfg{R}(p) \bfg{T}^{-1}(p)\jac{f}{y}(p))^{-1} \cdot (\jac{g}{x}(p) - \bfg{R}(p) \bfg{T}^{-1}(p)\jac{f}{x}(p)))]
    \Delta\bfg{x}
\end{equation}
which is in the form of \eqref{eq:singps}, where in this case $\xys\equiv \Delta \bfg x$ and 
\begin{equation}
  \label{eq:matrices:ode}
  \begin{aligned}
  \bfb{E}(p) &\equiv 
  \bfg I_n
  , \\ 
  \bfb{A}(p) &\equiv 
  \bfg{T}^{-1}(p) \cdot (\jac{f}{x}(p) - \jac{f}{y}(p)\cdot (\bfg{g}_y(p) - \bfg{R}(p) \bfg{T}^{-1}(p)\jac{f}{y}(p))^{-1} \cdot (\jac{g}{x}(p) - \bfg{R}(p) \bfg{T}^{-1}(p)\jac{f}{x}(p)))
  \end{aligned}
\end{equation}
Systems \eqref{eq:sidae:lin} and \eqref{eq:ode:lin} are dynamically equivalent, in the sense that they have the same $\nx$ finite eigenvalues.  Yet, \eqref{eq:sidae:lin} has additionally the infinite eigenvalue $s\rightarrow \infty$, with algebraic multiplicity $\ny$, see \cite{moebius}.
Moreover, the matrices that define \eqref{eq:sidae:lin} are sparse, whereas the state matrix in \eqref{eq:ode:lin} is dense, which is an important difference when it comes to numerical computations \cite{tzounas2020comparison}. In particular, for small systems, \eqref{eq:ode:lin} is preferred and the full set of its finite eigenvalues problem can be obtained efficiently, typically through QR factorization.  On the other hand, \eqref{eq:sidae:lin} is preferable for large systems.  A subset of the eigenvalues in this case can be found with appropriate sparse algorithms~\cite{tzounas2020comparison}.   
Apart from system size, an additional advantage of using \eqref{eq:sidae:lin} as opposed to \eqref{eq:ode:lin} is that the dimensions of $\bfb E$ and $\bfb A$ remain unaltered whenever a limit is reached or when a differential equation time constant is driven to zero.  This allows e.g., convenient detection of scenarios where an eigenvalue moves to infinity, by monitoring the rank of $\bfb{E}$.

%At this point we emphasize  a \ac{dae} formulation. Specifically, the dimensions of $\bfb P(s,p)$ do not change whenever a limit is reached or when a finite dynamic mode comes to an end (e.g., when a differential equation time constant is driven to zero), i.e., the value of $n$ reduces by $1$, while $m$ increases by the same amount.  In the \ac{dae} model this transition manifests itself as an eigenvalue moves toward infinity.  Monitoring the rank of $\bfb{E}$ allows to avoid spurious jumps to other modes.

Derivations in this paper are provided considering the general system form \eqref{eq:singps}.  Their specific expressions for \eqref{eq:sidae:lin} and \eqref{eq:ode:lin} can then be obtained as special cases in a straightforward way, i.e.,~by substituting the expressions of $\bfb{E}(p)$ and $\bfb{A}(p)$ from \eqref{eq:matrices:sidae}, \eqref{eq:matrices:ode}, respectively.  That said, we proceed to apply the Laplace transform to \eqref{eq:singps}:
\begin{equation}
\label{eq:dae:lin:lap}
[ s(p) \bfb{E}(p) - \bfb{A}(p) ]
\; \mathcal{L} \{ \xys \} = \bfb{E}(p)\xys(0) 
\end{equation}
where $s(p)$ is a complex frequency in the $S$-plane.  The polynomial matrix 
\begin{equation}
\label{eq:dae:pencil:sparse}
\bfb P(s,p) = s(p) \bfb E(p) - \bfb A(p)  
\end{equation}
is important in the study of \eqref{eq:singps} and is called the system's \textit{matrix pencil} \cite{milano2020eigenvalue,robust}.  In particular, every value of $s$ that when 
substituted to \eqref{eq:dae:pencil:sparse}
leads to a singular $\bfb P(s,p)$ is an eigenvalue of \eqref{eq:singps}, with the associated algebraic problem being:
\begin{align}
\label{eq:gep:r}
\bfb P(s,p) \; \bfg \phi(p) &= \bfg 0_{\nxy,1} \\
\label{eq:gep:l}
\bfg \psi(p) \; \bfb P(s,p) &= \bfg 0_{1,\nxy}
\end{align}
where every non-trivial vector $\bfg\phi(p)$ satisfying \eqref{eq:gep:r} is a right eigenvector; and every non-trivial vector $\bfg\psi(p)$ satisfying \eqref{eq:gep:l} is a left eigenvector.
The dimensions of $\bfb P(s,p)$ are $\nxy \times \nxy$, where the value of $r$ depends on whether $\bfb{E}$, $\bfb{A}$ are defined from \eqref{eq:matrices:sidae}, in which case $\nxy=n+m$, or \eqref{eq:matrices:ode}, in which case $\nxy=n$.

\section{Eigenvalue Tracking}
\label{sec:eig_track}

In this section, we describe the approach adopted to track targeted solutions of \eqref{eq:gep:r} as the
continuation parameter $p$ 
varies. We first focus on the tracking of a single eigenvalue and then address the problem of tracking multiple eigenvalues.

\subsection{Tracking a Single Eigenpair}

We start by differentiating \eqref{eq:gep:r} with respect to $p$:
\begin{equation}
\label{eq:gep:diff1}
\begin{aligned}
\dot{\bfb P}(s,p)
\; \bfg \phi(p) 
+ \bfb P(s,p)
\; \dot{\bfg \phi}(p) 
&= \bfg 0_{\nxy,1} 
\end{aligned}
\end{equation}
where:
\begin{equation}\label{eq:gep:diff2}
\begin{aligned}
\dot{\bfb P}(s,p)
&= \dot{s}(p) \bfb E(p) +
{s}(p) \dot{\bfb E}(p)
- \dot{\bfb A}(p) 
\end{aligned}
\end{equation}
and
$\dot s = \partial s / \partial p$,
$\dot{\bfb{\phi}} = \partial {\bfb{\phi}} / \partial p$,
$\dot{\bfb E} = \partial {\bfb E} / \partial p$, $\dot{\bfb A} = \partial {\bfb A} / \partial p$.

Equation \eqref{eq:gep:diff1} can be equivalently written as:
\begin{equation}\label{eq:gep:diff3}
\begin{aligned}
\bfb E \bfg \phi \dot{s} + [ s \bfb E - \bfb A ] \dot{\bfg \phi}  &= - [ {s} \dot{\bfb E} - \dot{\bfb A} ] \bfg \phi
\end{aligned}
\end{equation}
where the dependency on $p$ is omitted for simplicity. The dynamical system \eqref{eq:gep:diff3}, which describes 
the evolution a single eigenpair with respect to the parameter $p$, consists of $r$ equations and $r+1$ unknowns, namely $s$ and $\bfg\phi$.  To make this system well-posed, we additionally impose the following eigenvector normalization condition:
%
% \begin{equation}
% \label{eq:phi:norm}
% \begin{aligned}   
% \bfg{\phi}\T\bfb E\bfg{\phi} &= c
% \end{aligned}
% \end{equation}
%
\begin{equation}
\label{eq:phi:norm}
\begin{aligned}  
\bfg{\phi}\T\bfg{\phi} &= c
\end{aligned}
\end{equation}
where $c$ is a constant, e.g.,~$c=1$. 
Differentiation of \eqref{eq:phi:norm} gives \footnote{It is straightforward to show that $\dot{\bfg{\phi}}\T \bfg{\phi} = \bfg{\phi}\T \dot{\bfg{\phi}}$.}:
%
% \begin{equation}
% \label{eq:phi:norm_dif}
% \begin{aligned}   
% 2\bfg{\phi}\T\bfb E \dot{\bfg{\phi}} &= -\bfg{\phi}\T \dot{\bfb E} \bfg{\phi}
% \end{aligned}
% \end{equation}
%
\begin{equation}
\label{eq:phi:norm_dif}
\begin{aligned}
\bfg{\phi}\T \dot{\bfg{\phi}} &= 0
\end{aligned}
\end{equation}
%
%The derivation of \eqref{eq:phi:norm_dif} is given in the Appendix.

Combining \eqref{eq:gep:diff3}, \eqref{eq:phi:norm_dif}:
% %
% \begin{equation}
% \label{eq:gep:dyn}
% \begin{aligned}
% \begin{bmatrix}
% s \bfb E - \bfb A & \bfb E \bfg \phi \\
% 2{\bfg \phi}\T \bfb E & 0 
% \end{bmatrix}
% %
% \begin{bmatrix}
% \dot{\bfg \phi} \\
%  \dot{s} 
% \end{bmatrix}
% &= 
% \begin{bmatrix}
% - ( {s} \dot{\bfb E} - \dot{\bfb A} ) {\bfg \phi} \\
% - \bfg{\phi}\T \dot{\bfb E}{\bfg{\phi}}
% \end{bmatrix}
% \end{aligned}
% \end{equation}
% %
%
\begin{equation}
\label{eq:gep:dyn}
\begin{aligned}
\begin{bmatrix}
s \bfb E - \bfb A & \bfb E \bfg \phi \\
{\bfg \phi}\T & 0 
\end{bmatrix}
\begin{bmatrix}
\dot{\bfg \phi} \\
 \dot{s} 
\end{bmatrix}
&= 
\begin{bmatrix}
- ( {s} \dot{\bfb E} - \dot{\bfb A} ) {\bfg \phi} \\
0
\end{bmatrix}
\end{aligned}
\end{equation}
Splitting real and imaginary parts, 
i.e.,~$\bfg \phi = \bfg \phi_{\rm r} + \jj \bfg \phi_{\rm i} $, $s = s_{\rm r} + \jj s_{\rm i}$, we arrive to the following expression:
\begin{equation}
\label{eq:singlevector:sys}
\bfb M(\bfb y) \; \dot {\bfb y} = \bfg h (\bfb y)
\end{equation}
where $\bfb y = \bfb y(p) =(\bfg \phi_{\rm r}, \bfg \phi_{\rm i}, s_{\rm r}, s_{\rm i})$, with $\bfb y \in \mathbb{R}^{2\nxy + 2 } $; and:
% %
% \begin{align*}
% %\label{eq:singlevector:mat}
% \bfb M(\bfb y) &=
% \begin{bmatrix}
% s_{\rm r}\bfb{E}-\bfb{A} &-s_{\rm i} \bfb{E}
% & \bfb{E} \bfg \phi_{\rm r} & -\bfb{E}\bfg \phi_{\rm i} \\
% s_{\rm i} \bfb{E} & s_{\rm r} \bfb{E}-\bfb{A}
%         & \bfb{E}  \bfg \phi_{\rm i} & \bfb{E} \bfg\phi_{\rm r}  \\
%         2 \bfg\phi_{\rm r}\T{\bfb E} & -2 \bfg \phi_{\rm i}\T{\bfb E} &  0 &  0 \\
%         2 \bfg \phi_{\rm i}\T{\bfb E}  & 2 \bfg\phi_{\rm r}\T{\bfb E}  &  0 & 0
%     \end{bmatrix} \\
%      \bfg h (\bfb y) &= 
%     \begin{bmatrix}
%         \dot{\bfb{A}}\bfg\phi_{\rm r}-s_{\rm r}\Dot{\bfb{E}}\bfg\phi_{\rm r}
%         +s_{\rm i}\Dot{\bfb{E}}\bfg\phi_{\rm i}\\         \Dot{\bfb{A}}\bfg\phi_{\rm i}-s_{\rm i}\Dot{\bfb{E}}\bfg\phi_{\rm r}
%         -s_{\rm r}\Dot{\bfb{E}}\bfg\phi_{\rm i} \\ 
%         -\bfg\phi_{\rm r}\T\Dot{\bfb{E}}\bfg\phi_{\rm r} + \bfg\phi_{\rm i}\T\Dot{\bfb{E}}\bfg\phi_{\rm i}
%         \\ 
%         -\bfg\phi_{\rm r}\T\Dot{\bfb{E}}\bfg\phi_{\rm i} - \bfg\phi_{\rm i}\T\Dot{\bfb{E}}\bfg\phi_{\rm r}
%     \end{bmatrix}
% \end{align*}
% %
%
\begin{align*}
%\label{eq:singlevector:mat}
\bfb M(\bfb y) &=
\begin{bmatrix}
s_{\rm r}\bfb{E}-\bfb{A} &-s_{\rm i} \bfb{E}
& \bfb{E} \bfg \phi_{\rm r} & -\bfb{E}\bfg \phi_{\rm i} \\
s_{\rm i} \bfb{E} & s_{\rm r} \bfb{E}-\bfb{A}
        & \bfb{E}  \bfg \phi_{\rm i} & \bfb{E} \bfg\phi_{\rm r}  \\
       \bfg\phi_{\rm r}\T & -\bfg \phi_{\rm i}\T &  0 &  0 \\
       \bfg \phi_{\rm i}\T & \bfg\phi_{\rm r}\T  &  0 & 0
    \end{bmatrix} \\
     \bfg h (\bfb y) &= 
    \begin{bmatrix}
        \dot{\bfb{A}}\bfg\phi_{\rm r}-s_{\rm r}\Dot{\bfb{E}}\bfg\phi_{\rm r}
        +s_{\rm i}\Dot{\bfb{E}}\bfg\phi_{\rm i}\\         \Dot{\bfb{A}}\bfg\phi_{\rm i}-s_{\rm i}\Dot{\bfb{E}}\bfg\phi_{\rm r}
        -s_{\rm r}\Dot{\bfb{E}}\bfg\phi_{\rm i} \\ 
       0
        \\ 
        0
    \end{bmatrix}
\end{align*}
\eqref{eq:singlevector:sys} is a set of differential equations with state-dependent mass matrix $\bfb M(\bfb y)$.  Note that \eqref{eq:phi:norm_dif} is used instead
of \eqref{eq:phi:norm} because directly imposing the normalization as an algebraic constraint would result in a singular matrix $\bfb M(\bfb y)$.

Tracking is performed by numerically integrating \eqref{eq:singlevector:sys} over a parameter range
of interest $[p_{init},p_{fin}]$.  Given $p_{init}$, the corresponding initial condition $\bfb y(p_{init})$ for the state vector $\bfb y$ is obtained by solving once the eigenvalue problem \eqref{eq:gep:r}.  
In all subsequent steps, the system's equations and the matrices $\bfb A(p)$ and $\bfb E(p)$ are reconstructed.  Note that in cases where only the variation of parameters of dynamic devices like machines or controllers is of interest, 
the power flow solution remains unaltered, so tracking can proceed with the equilibrium updated from the steady-state DAEs at each step.
%and the power flow problem is solved when changes in parameters require the operating point to be updated.  
The matrix derivatives $\dot{\bfb E}$ and $\dot{\bfb A}$ in this paper are calculated numerically using first-order finite differences.

The integration proceeds iteratively until $p=p_{fin}$, as follows:
\begin{enumerate}
\item Given a step size $\Delta p$, the parameter is updated as $p_{k+1} = p_{k} + \Delta p$. 
\item The mass matrix $ \bfb M(\bfb y_k) $ and right-hand side $ \bfg h(\bfb y_k) $ are evaluated, and the state vector is updated using the chosen integration method. For example, using the forward Euler method yields:
\begin{equation}
  \label{eq:singlevector:fem}
  \bfb y_{k+1} = \bfb y_k + \Delta p \,\bfb M^{-1}(\bfb y_k) \,\bfg h (\bfb y_k)
\end{equation}
The product $\bfb M^{-1}(\bfb y_k) \,\bfg h (\bfb y_k)$ can be determined through \textit{LU decomposition} of $\bfb M(\bfb y_k)$.
\end{enumerate}

We note that, like any numerical integration-based scheme, the above process introduces a discretization error, which can be controlled by adapting the step size.  Moreover, the error can be effectively eliminated through the addition of a corrector step (e.g.,~using Newton's method), at an additional computational cost.  This is further discussed in Section~\ref{sec:case}.

%Like every numerical integration method, forward Euler method introduces a discretization error, which depends on both the method and the integration step. Regardless, using the result of \eqref{eq:singlevector:fem} as initial guess, \eqref{eq:singlevector:sys} can be combined with a corrector sub-routine in order to achieve the desired accuracy, at the expense of computational cost. Alternative approaches (e.g., see~\cite{li2017eigenvalue}) that use the previous solution as initial guess between iteration -- thus dismissing the predictor step -- can lead the solution to diverging or converging to the wrong value. This issue is later discussed in the case study.

\subsection{Tracking Multiple Eigenpairs}
\label{mvt}

We extend the above approach to the case of multiple eigenvalues. 
Let $\bfg{\Phi} \in \mathbb{C}^{\nxy \times \kappa}$ be the matrix whose columns are linearly independent right eigenvectors of $\kappa$ finite 
eigenvalues of \eqref{eq:dae:pencil:sparse}, i.e.,~$\bfg{\Phi} = [ \bfg{\mathbf{\phi}}_1 \, \bfg{\mathbf{\phi}}_2 \, \ldots \, \bfg{\phi}_\kappa ]$, where $1 \leq \kappa \leq \nx$. 
Then, \eqref{eq:gep:r} can be generalized as:
\begin{equation}
\label{eq:gep:mul}
    \bfb{A}(p) \bfg{\Phi}(p) - \bfb E(p) \bfg \Phi(p) \bfg{\Lambda}(p) = \bfg 0_{\nxy,\kappa} 
\end{equation}
where $\bfg{\Lambda}$ is diagonal matrix containing the $\kappa$ finite eigenvalues of interest. Differentiation with respect to 
the continuation parameter $p$ 
gives:
\begin{equation}\label{eq:mvt:dif}
    \begin{aligned}
        {\bfb{A}} \dot{\bfg{\Phi}} 
        + \dot{\bfb{A}} {\bfg{\Phi}}
        - \dot{\bfb{E}} {\bfg{\Phi}} \bfg{\Lambda}
        - {\bfb{E}} \dot{\bfg{\Phi}} \bfg{\Lambda}
        - {\bfb{E}} {\bfg{\Phi}} \dot{\bfg\Lambda}
         &= \bfg 0_{\nxy,\kappa} \\
    \end{aligned}
\end{equation}
or, equivalently:
\begin{equation}
\label{eq:gep:dyn2}
\begin{aligned}
\begin{bmatrix}
\bfb A & -\bfb E {\bfg{\Phi}} \\
\end{bmatrix}
\begin{bmatrix}
\Dot{\bfg{\Phi}} \\
\dot{\bfg\Lambda}
\end{bmatrix}
-
\begin{bmatrix}
\bfb E & \bfg 0 \\
\end{bmatrix}
\begin{bmatrix}
\Dot{\bfg{\Phi}} \\
\dot{\bfg\Lambda}
\end{bmatrix}
%\begin{bmatrix}
{\bfg\Lambda}
&= 
%\begin{bmatrix}
\dot{\bfb{E}} {\bfg{\Phi}}\bfg{\Lambda}
- \dot{\bfb{A}} {\bfg{\Phi}}\\
\end{aligned}
\end{equation}
Similar to \eqref{eq:gep:diff3}, suitable normalization constraints can be introduced to make \eqref{eq:gep:dyn2} well posed. In the multiple-eigenpair case, this leads to a generalized Sylvester equation of the form:
\begin{equation}\label{eq:sylvester} 
\bfg{M} \bfg{X} + \bfg{G} \bfg{X} \bfg{N} = \bfg{F}
\end{equation}
where $\bfg{X} = (\bfg{\Phi}, \bfg{\Lambda} )$.   To the best of our knowledge, efficient solvers for the numerical solution of \eqref{eq:sylvester} in its general form are not available. In the special case where $\bfg{G}$ is the identity matrix, the Bartels-Stewart algorithm can be used, see \cite{BOUHAMIDI2008687, ZHOU2008200, JIN2020, SIMONCINI2016}. However, the algorithm does not scale well with system size and thus appears impractical for large-scale power system models.
Given these limitations, in this paper we track multiple eigenvalues by applying the single-eigenpair formulation \eqref{eq:singlevector:sys} separately to each mode of interest. The tracking process is repeated independently for every eigenpair and is therefore naturally parallelizable.

\section{Case Study}
\label{sec:case}

This section presents simulation results on the eigenvalue tracking approach described in Section~\ref{sec:eig_track}. The results in Sections~\ref{case:39bus} and \ref{case:39bus:mod} are based on the IEEE 39-bus system, detailed data of which can be found in \cite{web:39bus}.  Then, Section~\ref{case:aiits} considers a real-world size dynamic model of the \ac{aiits}. In all cases, eigenvalue trajectories are compared against those computed via repeated, dense QR factorizations, which serve as the reference.  All simulations are carried out with the power system software tool Dome \cite{dome}.  The version of the software employed in this paper relies on LAPACK~3.10.0 for QR factorization
and KLU~5.10.1 for LU factorization.

\subsection{IEEE 39-Bus System}
\label{case:39bus}

\begin{figure}[ht!]
  \centering
\includegraphics[width=0.8\linewidth]{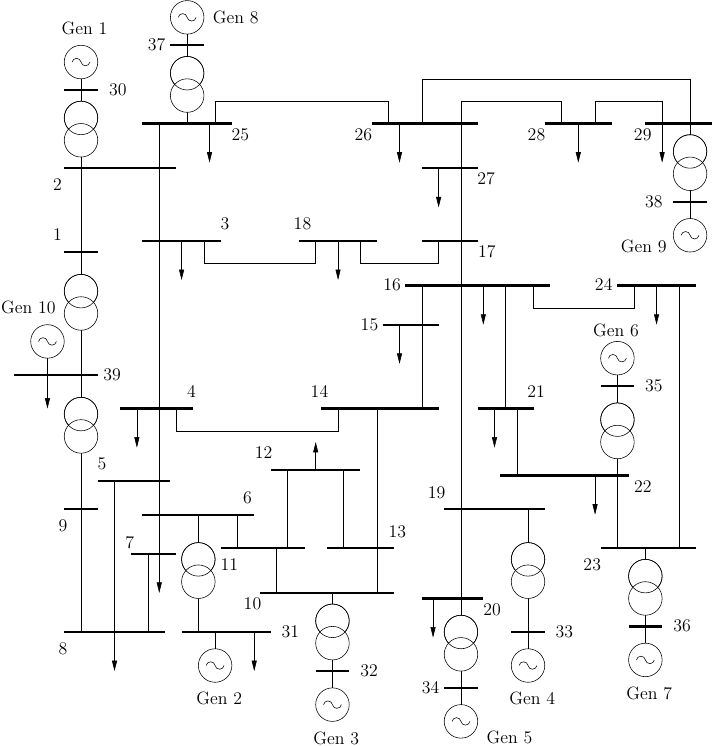}
  \caption{39-bus system: single-line diagram.} 
  \label{fig:39b_SLD}
\end{figure}

The results of this section are based on the IEEE 39-bus system 
(see Fig.~\ref{fig:39b_SLD}).  
The system comprises 10 \acp{sm}, represented by fourth-order models.  Machines are equipped with automatic voltage regulators, power system stabilizers, and \acp{tg}.   For the purposes of this study, the original system has been modified to reduce by a factor of ten the inertia constant of the \ac{sm} connected to bus~39.

\begin{figure}[ht!]
  \centering
   \begin{subfigure}[t]{\columnwidth}
    \centering      \includegraphics[width=0.6\linewidth]{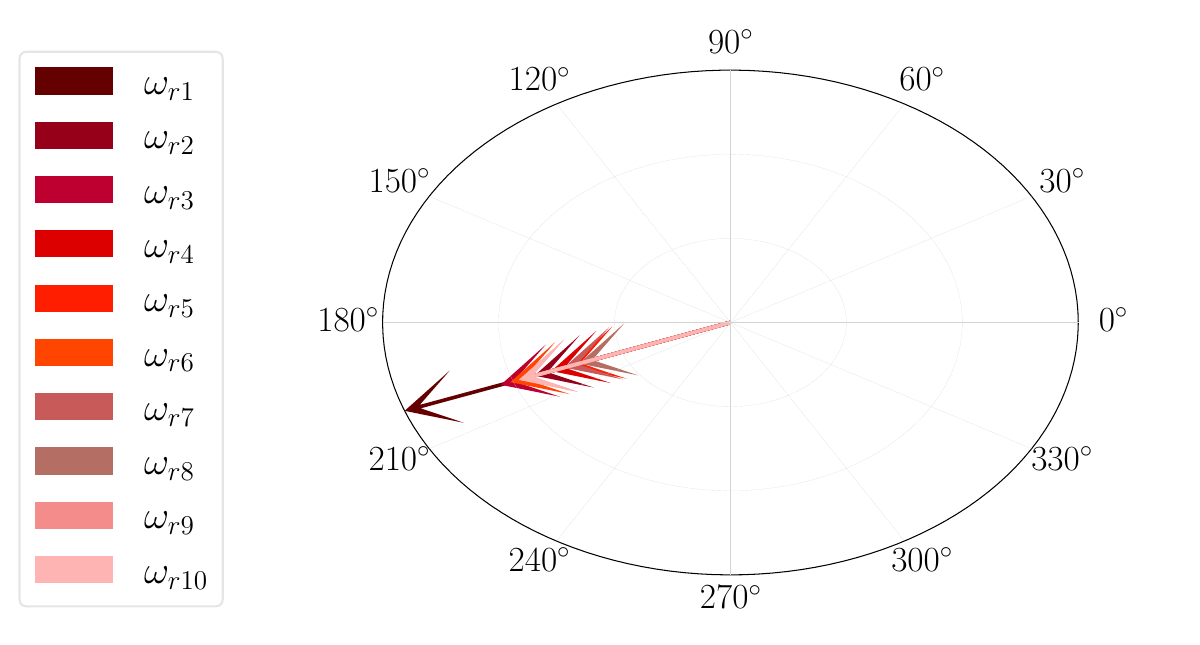}
      \caption{Frequency regulation~(FR) mode.}
      \label{fig:PF_39b_TG_R_002_102}
   \end{subfigure}
   \begin{subfigure}[t]{\columnwidth}
    \centering  
    \includegraphics[width=0.6\linewidth]{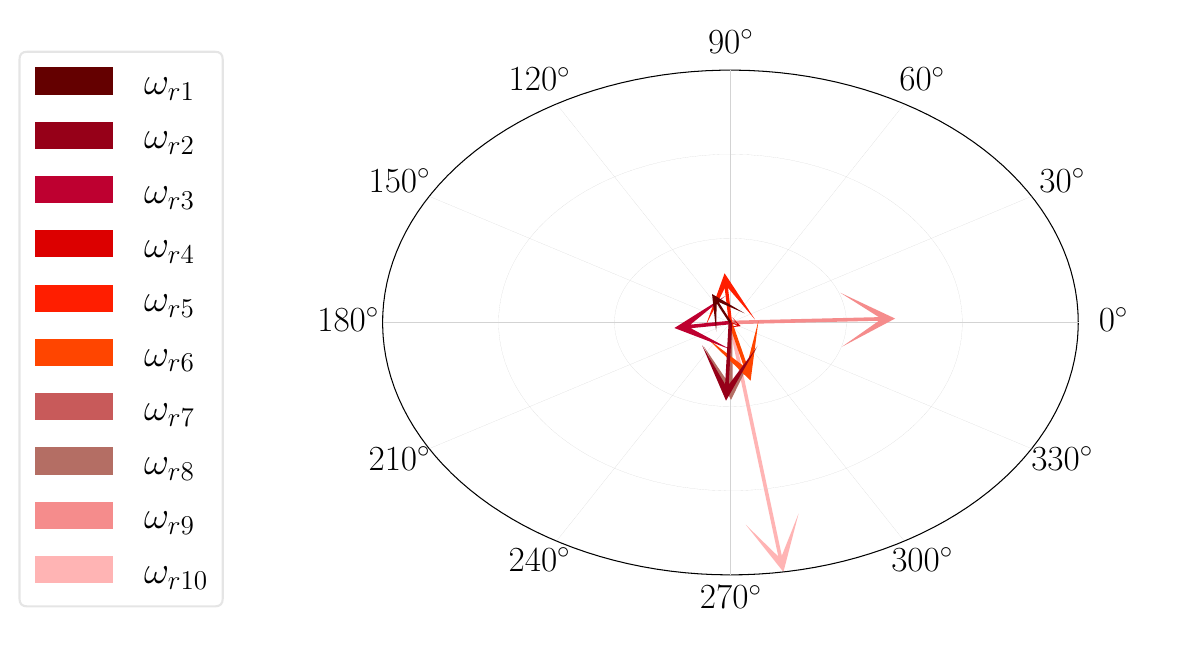}
      \caption{Local electromechanical mode of SM at bus~30.}
      \label{fig:PF_39b_TG_R_005_120}
  \end{subfigure}
\caption{39-bus system: \ac{sm} rotor speed participation factors.}
\label{fig:PF_39b_TG_R_002}
\end{figure}

\begin{figure}[ht!]
  \centering
\includegraphics[width=0.8\linewidth]{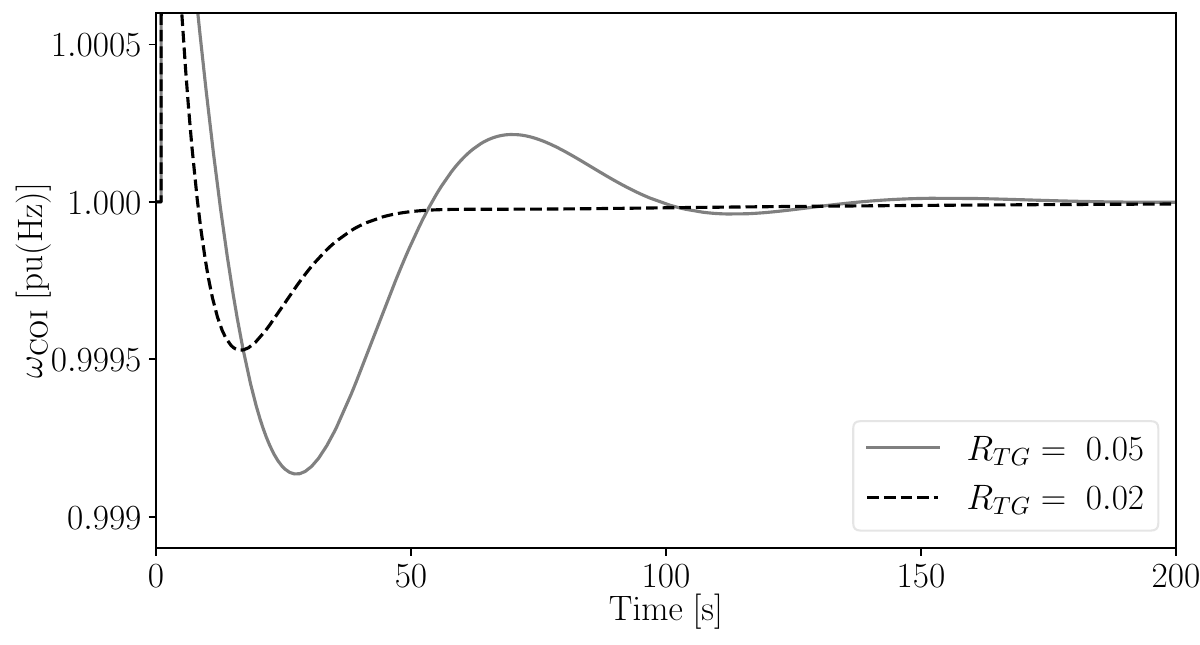}
  \caption{39-bus system: Effect of \ac{tg} droop on $\omega_{\rm COI}$.} % of \ac{sm}~$8$.}
  \label{fig:39b_TG_R_TDS}
\end{figure}

We focus on tracking the system's \acf{fr} mode.  This mode emerges with the inclusion of frequency controllers in the system and reflects a coherent response to frequency deviations across the network.  Due to its global coherence and the relatively slow response of conventional turbine governors, this mode has also been referred to in the literature as the \textit{common low-frequency mode}. \ac{fr} modes have drawn interest due to their relevance in understanding potential oscillatory behaviors associated with closed-loop frequency control dynamics, see~\cite{Bernal_2016, Moeini_2016,2023enhancing}. 
At the base case operating point of the examined system, the \ac{fr} mode is represented by the complex pair of eigenvalues $-0.03533 \pm \jj 0.07357$, which has natural frequency $0.01$~Hz and damping ratio $43.3$\%.  As shown in Fig.~\ref{fig:PF_39b_TG_R_002_102}, the mode is characterized by large, phased-aligned participation factors associated with all \ac{sm} rotor speeds, indicating its system-wide nature.  In this figure, a droop constant $R_{TG} = 0.05$ is used for all \acp{tg}.  For the sake of comparison,  Fig.~\ref{fig:PF_39b_TG_R_005_120} shows the participation factors for an electromechanical mode, which is local to the \ac{sm} at bus~30.  A time-domain simulation of the system further illustrates the behavior of the identified \ac{fr} mode. We consider a three-phase fault at bus~6, occurring at $t = 1.0$~s and cleared after 80~ms by opening the line that connects buses 5~and~6. The response of the \ac{coi} frequency is shown in Fig.~\ref{fig:39b_TG_R_TDS} and confirms the presence of a low-frequency oscillation around 0.01~Hz. As the droop constant $R_{TG}$ is reduced the mode becomes faster, which is as expected.  

\begin{figure}[ht!]
  \centering
  \includegraphics[width=0.8\linewidth]{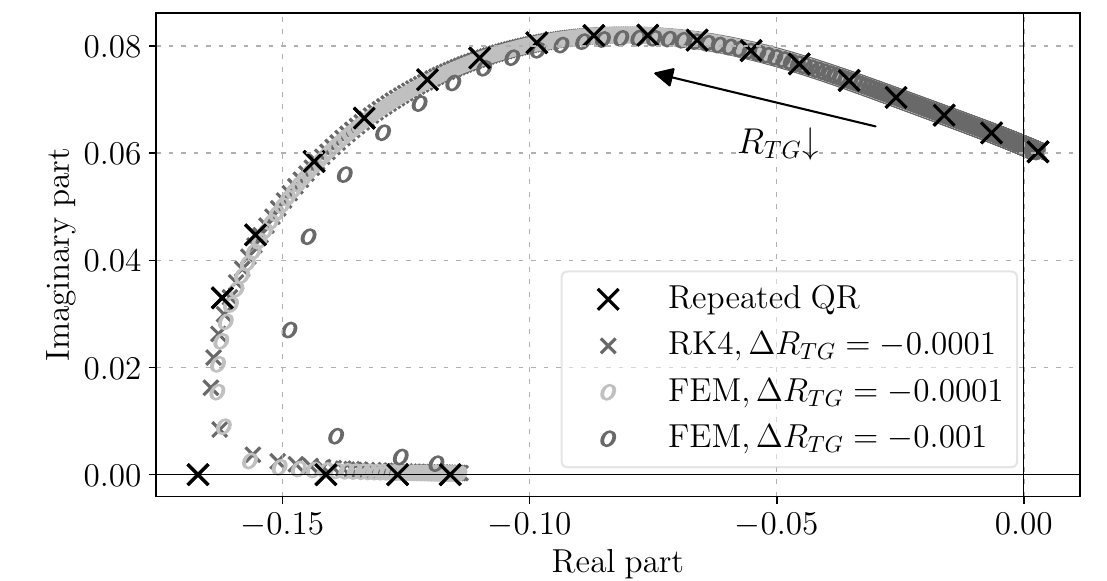}
  \caption{39-bus system, \ac{fr} mode: Effect of varying $R_{TG} \in [0.01,0.2]$.}
  \label{fig:39b_TG_R}
\end{figure}

In the following, we employ the eigenvalue tracking method \eqref{eq:singlevector:sys}
to trace the \ac{fr} mode under variations of key system parameters.  We start by examining how the corresponding complex pair of eigenvalues evolves as the droop constant $R_{TG}$ of the \acp{tg} is varied.  Starting from an initial large value of $R_{TG}=0.2$, 
%(at which the operating point is unstable)
the trajectory is tracked as the parameter decreases to $0.02$.  In Fig.~\ref{fig:39b_TG_R}, the results obtained via repeated QR factorizations serve as a reference.  Performance of \eqref{eq:singlevector:sys} is then illustrated for two numerical integration methods: \ac{fem} and \ac{rk4}. 
To this end, no corrector step is employed in this case.
Both methods accurately follow the mode trajectory when a sufficiently small parameter step is used. For the remainder of this case study, we use \ac{fem} due to its simplicity, although any standard integration method is applicable. Computational efficiency, which is of particular interest in large-scale systems, is discussed in Section~\ref{case:aiits}.

Figure~\ref{fig:39b_TG_R} is selected particularly because it showcases the behavior of \eqref{eq:singlevector:sys} when it comes to capturing the trajectories of eigenvalues that go through a change of their spectral structure. Specifically, as $R_{TG}$ decreases, a critical point is reached at which the complex pair of eigenvalues representing the \ac{fr} mode becomes real. At this point, the eigenvalue becomes defective with algebraic multiplicity $2$ and geometric multiplicity $1$; that is, the system lacks a complete set of linearly independent eigenvectors as.  This point corresponds to a \textit{simple quadratic fold}. It occurs at a specific parameter value $p$ for which the partial derivative $\partial s/\partial p$ of the traced eigenvalue does not exist, and the associated eigenvector becomes discontinuous \cite{sun1990multiple}.

As $R_{TG}$ is further decreased, the defective eigenvalue in turn splits into two distinct real eigenvalues.  Thereon, the tracking method inherently continues along one of the two resulting branches, determined by the numerical evolution of \eqref{eq:singlevector:sys}. 
In many cases, studying the second branch may not be of interest as, at this point, both eigenvalues are perfectly damped.  However, if the second branch is indeed of interest, we recommend that in practice the tracking process is reinitialized after the spectral splitting by recomputing an eigendecomposition of the system's pencil and extracting the eigenpair corresponding to that branch.  The numerical and computational implications of this approach are further discussed later in this case study.

From Fig.~\ref{fig:39b_TG_R}, we also observe that a step size of $\Delta R_{TG} =-0.0001$ yields accurate tracking of the \ac{fr} mode.  Naturally, the choice of integration step size depends on the sensitivity of the eigenvalue to the parameter being varied.  As an example, the same figure shows the impact of increasing the step size tenfold.  Table~\ref{tab:39b_R_TG} 
reports both the relative error in the eigenvalue and the error in its damping ratio for the two step sizes, highlighting the method's accuracy in capturing the mode's evolution.  In practice, a key consideration for step-size selection is the Euclidean distance between the tracked eigenvalue at consecutive steps, i.e., between $s_k$ and $s_{k+1}$ computed via \eqref{eq:singlevector:sys}.
This metric can also inform adaptive step-size control strategies, an aspect further discussed in Section~\ref{case:aiits}.

\begin{table}[!ht]
\caption{39-bus system: Tracking accuracy as $R_{TG}$ is varied.}
\label{tab:39b_R_TG}
\centering
\begin{tabular}{c|cc|cc}
\toprule\toprule
\centering
 & \multicolumn{2}{c}{\textbf{$\Delta R_{TG} = 0.0001$}} & \multicolumn{2}{c}{\textbf{$\Delta R_{TG} = 0.001$}}\\
\textbf{$R_{TG}$}  & {Relative error} & {$\Delta \zeta [\%]$} & {Relative error} & {$\Delta \zeta [\%]$}\\
\midrule
$0.15$ & $1.14 \times 10^{-6}$ & $0.00105$ & $0.00011$ & $0.00944$\\
$0.10$ & $5.52 \times 10^{-5}$ & $0.00543$ & $0.00055$ & $0.04874$\\
$0.05$ & $0.00034$ & $0.02345$ & $0.0034$ & $0.21196$\\
$0.01$ & $0.0046$ & $0.00012$ & $0.04875$ & $0.01472$\\
\bottomrule\bottomrule
\end{tabular}
\end{table}

\begin{figure}[ht!]
  \centering
  \includegraphics[width=0.8\linewidth]{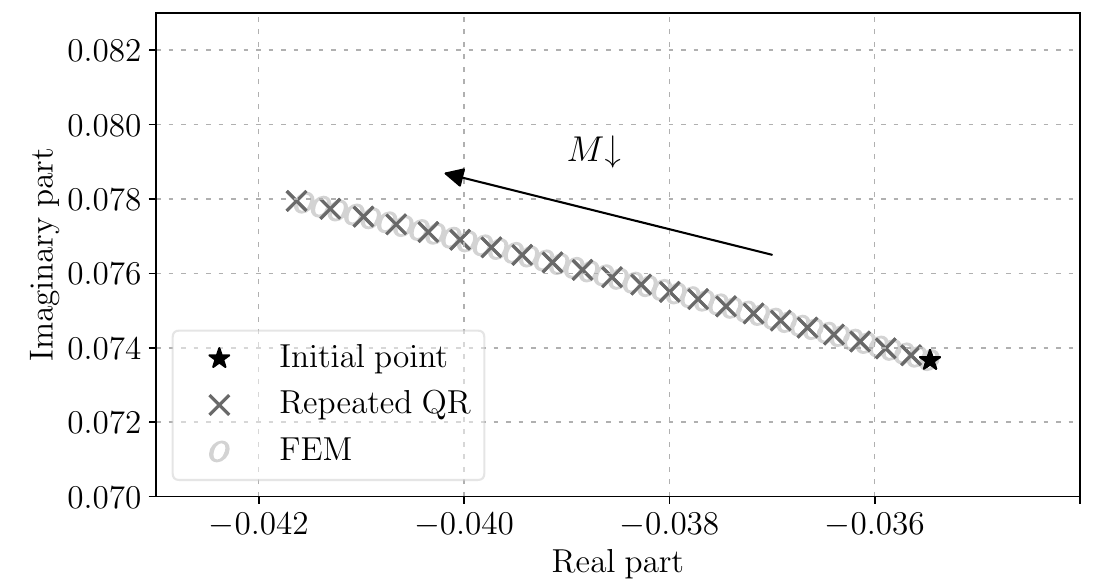}
  \caption{39-bus system, \ac{fr} mode: Effect of gradually reducing up to $90\%$ the inertia of the \ac{sm} at bus 39.}
  \label{fig:39b_M_syn1}
\end{figure}

Figure~\ref{fig:39b_M_syn1} shows the performance of \eqref{eq:singlevector:sys}
when tracking the \ac{fr} mode during a gradual decrease of up to $90 \%$ in the inertia constant $M$ of the system's largest \ac{sm}, which is connected to bus~39.  This decrease in system inertia increases the speed of the \ac{fr} mode. Notably, high tracking accuracy is maintained even for a large step of $\Delta M = -1$~pu~[s].   

\begin{figure}[ht!]
  \centering
  \includegraphics[width=0.8\linewidth]{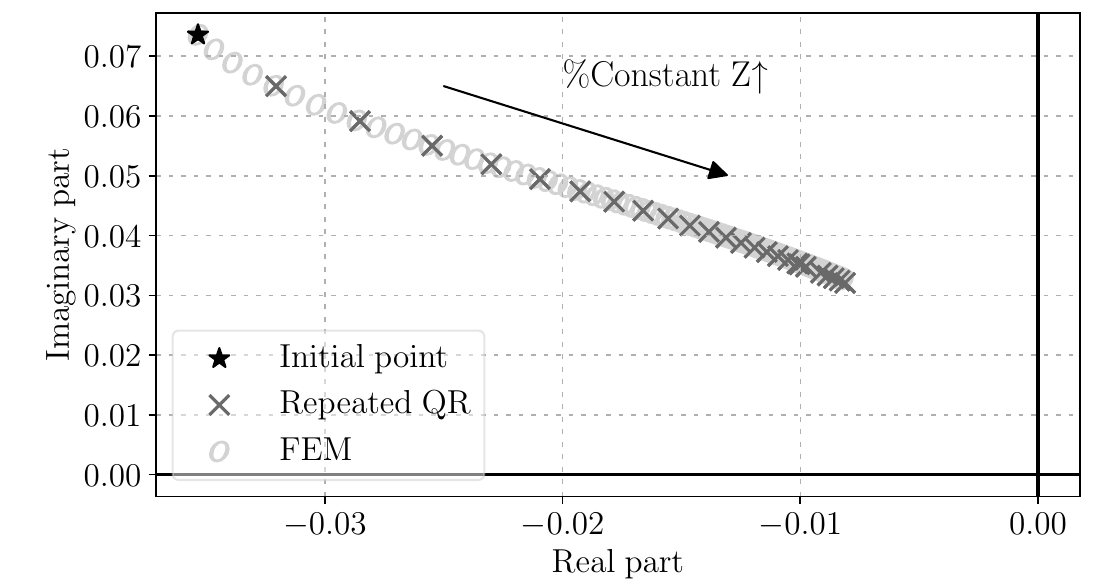}
  \caption{39-bus system, \ac{fr} mode: Effect of varying the share of constant impedance loads of the system from $0$ to $100\%$.}
  \label{fig:39b_Z_P}
\end{figure}

We next examine the accuracy of \eqref{eq:singlevector:sys} under changes in the system's load model composition. Specifically, loads are modeled as a mix of constant power and constant impedance ($\rm Z$) types.  The loads composition also alters the power flow solution, along with the operating point, thus the solution needs to be updated before each iteration of the tracking in this case.
Starting from a system with all loads being of constant power consumption, the share of constant impedance loads is gradually increased in $1\%$ steps up to $100\%$. As shown in Fig.~\ref{fig:39b_Z_P}, with the increase of this share, the speed of the \ac{fr} mode drops from $0.013$~Hz to $0.005$~Hz, while its damping reduces, specifically by $20\%$.  

\subsection{Impact of Inverter-Based DERs}
\label{case:39bus:mod}

This section considers a modified version of the IEEE 39-bus system, in which half of the \acp{sm} are replaced by inverter-based \acp{der}.  Specifically, the \acp{sm} at buses 30, 34, 35, 36 and 37 are replaced by \acp{der} of the same capacity.  
Each \ac{der} is represented by an inner control loop that regulates the $d$ and $q$ components of the current in the $dq$ reference frame, along with two outer loops for primary frequency and voltage control \cite{ORTEGA201837}.  The focus is again on tracking the \ac{fr} mode via \eqref{eq:singlevector:sys}.

\begin{figure}[ht!]
\centering
  \includegraphics[width=0.8\linewidth]{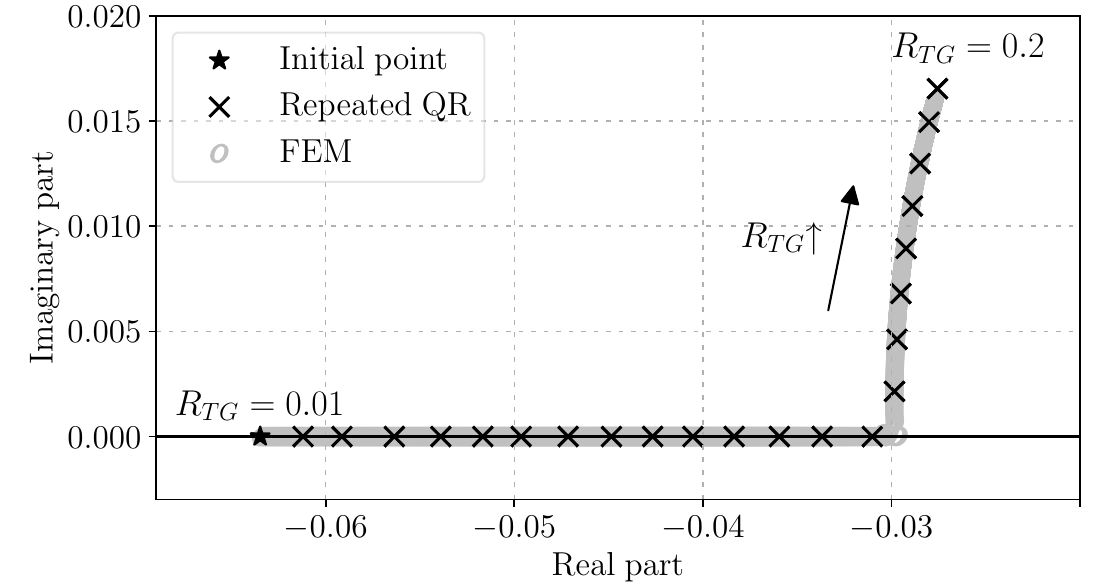}
  \caption{Modified 39-bus system, \ac{fr} mode: Effect of varying \ac{tg} droops in the range $[0.01,0.2]$.}
  \label{fig:mod39b_TG_R}
\end{figure}

We start by gradually decreasing, from $0.2$ to $0.01$, the \ac{tg} droop constants of the five remaining \acp{sm}.  This variation leads to faster frequency control response and yields conclusions similar to those made for the original system.  As shown in Fig.~\ref{fig:mod39b_TG_R}, decreasing $R_{TG}$ results in a gradual increase of the \ac{fr} mode's damping, reaching $100\%$ at which point the complex pair of eigenvalues becomes real.  Equation \eqref{eq:singlevector:sys} can accurately track the eigenvalue of interest when the latter remains non-defective throughout the parameter variation.  However, the transition from a complex-conjugate pair to two real eigenvalues passes through a defective eigenvalue.  In this case, \eqref{eq:singlevector:sys} inherently follows one of the resulting real eigenvalues, as shown in Fig.~\ref{fig:mod39b_TG_R}, where the imaginary part diminishes smoothly along the tracked trajectory.

When tracking is instead initialized from a real eigenvalue and the parameter is increased, the continuation-based formulation is unable to produce a complex pair at the point where the two real eigenvalues merge.  To address this, a small imaginary part is introduced to the initial condition.  Specifically, a term $\jj \epsilon$ is added to the initial eigenvalue $s_{\rm i}(p_{init})$ and to corresponding right eigenvector $\bfg \phi_{\rm i}(p_{init})$.  In this way, the method can continue consistently through the transition and track the eigenvalue in the full range $R_{TG}\in [0.01 , 0.2]$.

\begin{figure}[ht!]
  \centering
   \begin{subfigure}[t]{\columnwidth}
    \centering  \includegraphics[width=0.8\linewidth]{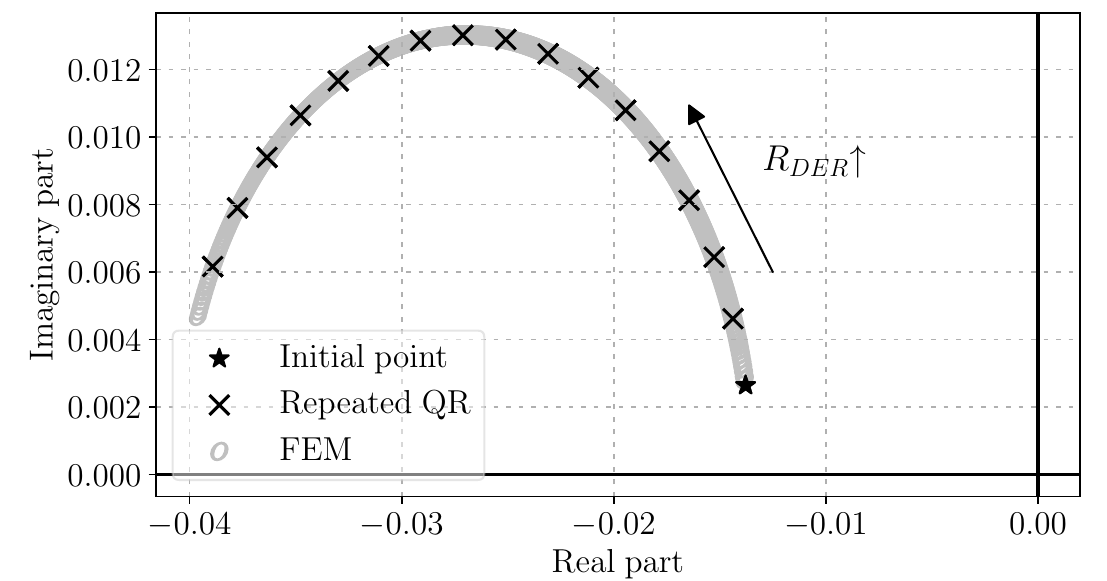}
    \caption{Increasing the droop constant $R_{DER}$ from $0.01$ to $0.07$.}
    \label{fig:mod39b_DER_FREQ_R}
   \end{subfigure}
   \begin{subfigure}[t]{\columnwidth}
    \centering  \includegraphics[width=0.8\linewidth]{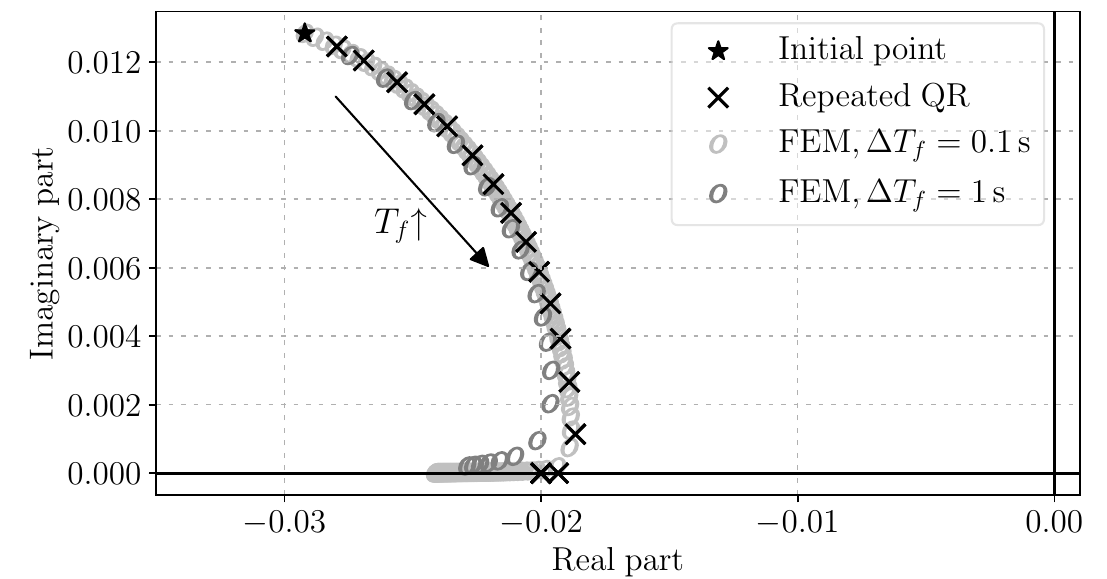}
    \caption{Increasing the droop time constant $T_{f}$ from $0.1$ to $99.0$~s.}
    \label{fig:mod39b_DER_FREQ_Tf}
  \end{subfigure}
  \caption{Modified 39-bus system, \ac{fr} mode: Effect of varying the droop parameters of the \ac{der} frequency controllers.}
  \label{fig:mod39b_DER_FREQ}
\end{figure}

We next examine how the \ac{fr} mode changes as the droop constants $R_{DER}$ of the \ac{der} frequency controllers are varied.  As shown in Fig.~\ref{fig:mod39b_DER_FREQ_R}, the eigenvalue associated with the \ac{fr} mode moves progressively leftward in the complex plane as $R_{DER}$ increases.  Moreover, the damping ratio exhibits a non-monotonic trend: it decreases until reaching a minimum value of $87.35\%$ 
at $R_{DER} = 0.027$, after which it begins to increase.  This behavior is explained by considering the relative roles of \acp{der} and \acp{sm} in frequency regulation.  With the \ac{tg} droop constants of \acp{sm} fixed at $R_{TG}=0.05$, frequency regulation is initially dominated by the \acp{der}. As $R_{DER}$ increases, their frequency response weakens, and the \acp{sm} play an increasing role in shaping the mode's structure.

Varying the time constant $T_f$ of the low-pass filter of the \ac{der} droop controllers has the opposite effect, as shown in Fig.~\ref{fig:mod39b_DER_FREQ_Tf}.  
As $T_f$ increases, the controller response becomes slower, eventually suppressing the \ac{fr} mode as the associated complex eigenvalues become real. The accuracy of \eqref{eq:singlevector:sys} during this transition is summarized in Table~\ref{tab:39mod_T_f}. 
We note that the results in Table~\ref{tab:39mod_T_f}  Fig.~\ref{fig:mod39b_DER_FREQ} do not make use of the corrector step.

\begin{table}[!ht]
\caption{Modified 39-bus system: FR mode tracking as $T_f$ varies.}
\label{tab:39mod_T_f}
\centering
\begin{tabular}{c|cc|cc}
\toprule
\centering
{}
 & \multicolumn{2}{c}{$\Delta T_{f} = 0.1$~s} & \multicolumn{2}{c}{$\Delta T_{f} = 1$~s}\\
 $T_{f}$~[s] & {Relative error} & \textbf{$\Delta \zeta \%$} & {Relative error} & \textbf{$\Delta \zeta \%$}\\
\midrule
$0.1$ & $0$ & $0$ & $0$ & $0$ \\
%\midrule
$1$ & 
%$7.18 \times 10^{-5}$ 
$0.00007$
& $0.00283$ & $0.00229$ & $0.02231$ \\
%\midrule
$10$ & $0.00059$ & $0.02183$ & $0.00573$ & $0.22536$\\
%\midrule
$20$ & $0.00105$ & $0.02591$ & $0.01001$ & $0.26364$\\
%\midrule
$50$ & $0.00649$ & 
\!\!\!\!$-0.00123$ 
& $0.06459$ & 
\!\!\!\!$-0.20870$ \\
$99$ & $0.20693$ & 
%$-1.51 \times 10^{-5}$ & $0.20469$ 
\!\!\!\!$-0.00002$
& \!\! $0.20469$ & \!\!\!\! $-0.00154$\\
\bottomrule
\end{tabular}
\end{table}

\begin{figure}[ht!]
  \centering
  \includegraphics[width=0.8\linewidth]{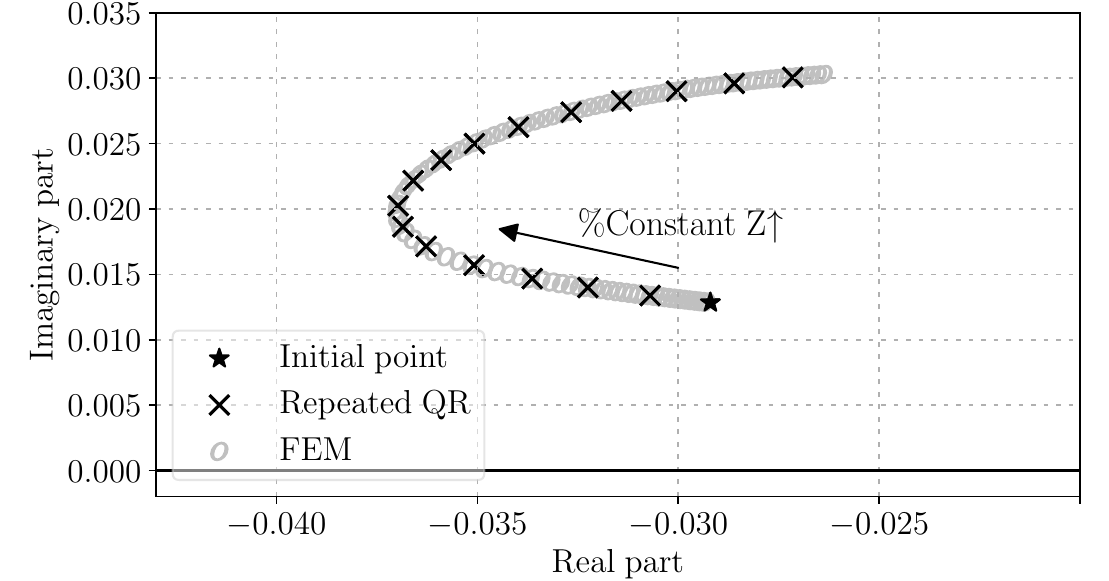}
  \caption{Modified 39-bus system, \ac{fr} mode: Effect of varying the percentage of constant impedance loads from $0$ to $100\%$.}
  \label{fig:mod39b_Z_P}
\end{figure}

We then study the effect of the load model composition.  The first aspect concerns 
the effect of increasing the share of constant impedance loads, which is examined in Fig.~\ref{fig:mod39b_Z_P}. As this share increases, the damping of the \ac{fr} mode decreases from $91.5 \%$ to $65.5 \%$.  Meanwhile, the mode's natural frequency reaches a maximum of $0.007$~Hz when constant impedance loads are about $50 \%$ of the total.

\begin{figure}[ht!]
  \centering
  \begin{subfigure}[t]{\columnwidth}
    \centering  
  \includegraphics[width=0.8\linewidth]{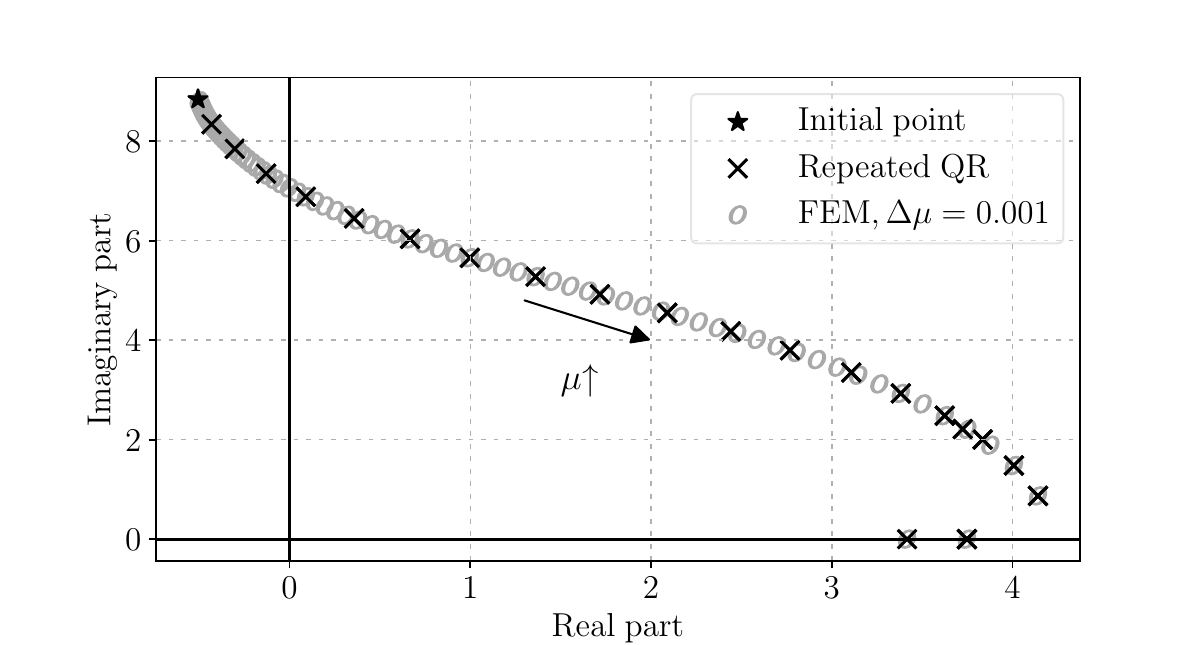}
  \caption{Local electromechanical mode of the SM at bus~39.}
  \label{fig:mod39b_QR_LEM}
  \end{subfigure}
   \begin{subfigure}[t]{\columnwidth}
    \centering  \includegraphics[width=0.8\linewidth]{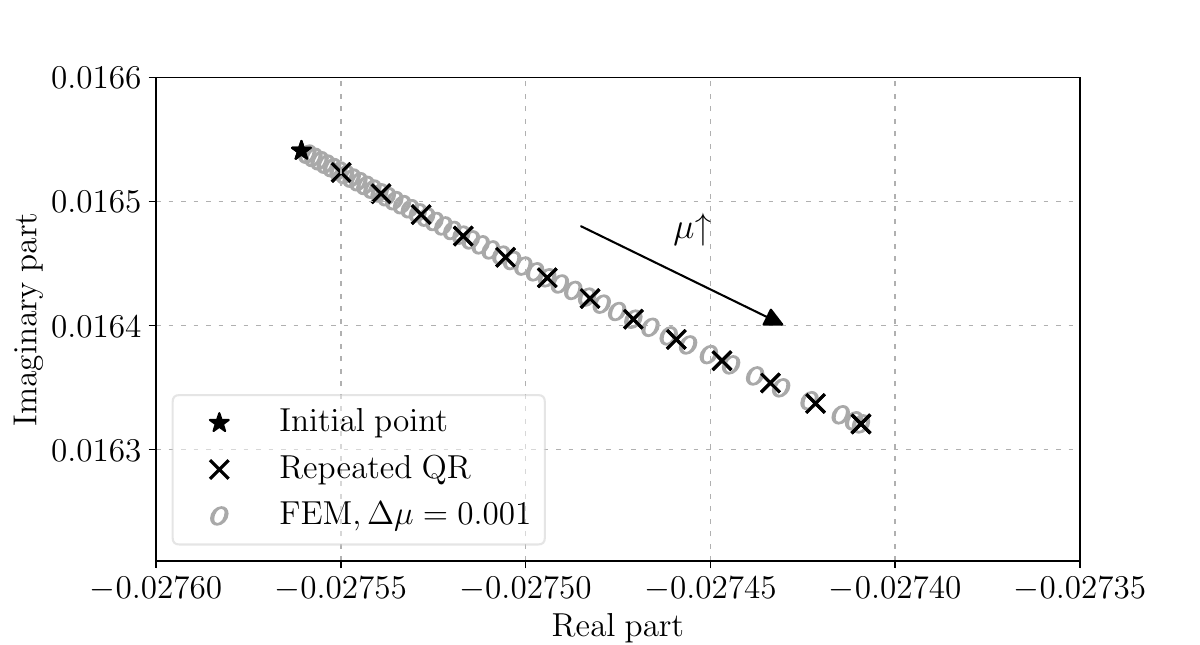}
    \caption{FR mode.}
    \label{fig:mod39b_QR_FRM}
  \end{subfigure}
  \caption{Modified 39-bus system: Effect of increasing the power of the system's loads.}
  \label{fig:mod39b_QR}
\end{figure}

We then demonstrate how \eqref{eq:singlevector:sys} handles the simultaneous increase of the active and reactive power of the system's loads.   To this end, we use a loading parameter $\mu$, and starting from $\mu=0$, the power of the loads are increased by a factor of $1+\mu$, up to $\mu=0.23$, similar to a continuation power flow process \cite{ajjarapu2002continuation, milano2010continuation}.   By solving the power flow problem at each step, we update the values of generated power and, subsequently,
matrices $\bfb E$ and $\bfb A$. 
Figure~\ref{fig:mod39b_QR} shows the tracking results as the system's gets closer to its maximum loadability limit.
%This scenario resembles the continuation power flow technique, used to identify the maximum loadability limit of power systems (see~), on which a saddle node bifurcation causes voltage collapse, and the power flow has no solution.
%More precisely, the magnitude of the loads' power  -- along with the power generated -- are gradually increased, causing a shift to the operating point and the solution of the power flow problem.  
%The theoretical maximum loading point leads to voltage collapse and offers a means to test the tracking close to a \textit{saddle node bifurcation}.
%In our case, given that this limit is not reached, we can trace eigenvalues near the bifurcation point, 
In particular,
Fig.~\ref{fig:mod39b_QR_LEM} shows the trajectory of the local electromechanical mode of the SM at bus~39, initially located at $-0.50650 \pm \jj 8.84058$. The mode is driven into instability, with a transition from a complex pair to two real and positive eigenvalues taking place eventually.
%For the greatest part of this trajectory, the system's spectrum corresponds to unstable equilibrium points. Therefore, to smoothly increase the system's loads is not likely to lead the system dynamics to the states described by the rightmost points of Fig.~\ref{fig:mod39b_QR}, despite the accuracy of \eqref{eq:singlevector:sys}.
The effects on \ac{fr} mode are negligible, as shown in Fig.~\ref{fig:mod39b_QR_FRM}.

\begin{figure}[ht!]
  \centering
  \includegraphics[width=0.8\linewidth]{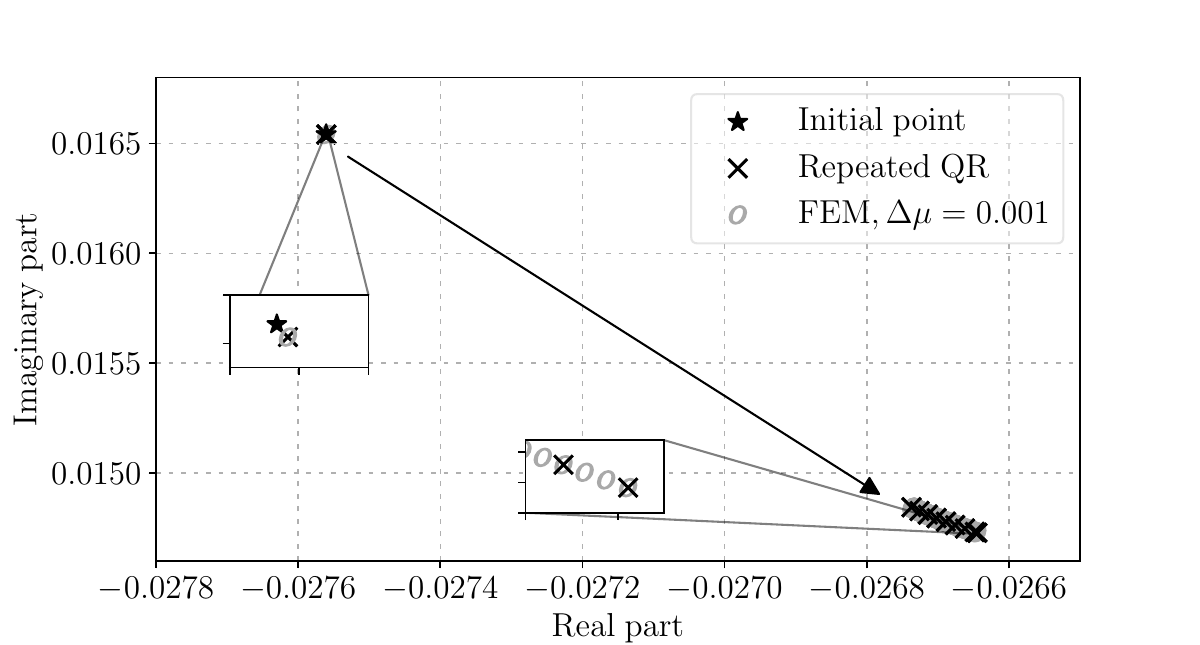}
  \caption{Modified 39-bus system, \ac{fr} mode: Effect of reaching a \ac{tg} active power limit.}
  \label{fig:mod39b_QR_TG_jump}
\end{figure}

Finally, we examine tracking performance in the event of a control saturation.  To this end, we lower the maximum active power limit of the system's \acp{tg} to values close to the power generated by the corresponding \acp{sm}.  
Figure~\ref{fig:mod39b_QR_TG_jump} shows the effect on the \ac{fr} mode, when the loading parameter $\mu$ increases in the same range as in Fig.~\ref{fig:mod39b_QR}.  For $\mu=0.002$ the active power limit of the \ac{tg} connected to bus~31 is reached, and the eigenvalue trajectory becomes discontinuous as a jump is observed. Nonetheless, \eqref{eq:singlevector:sys} converges to the correct point and completes the tracking in the designated range of $\mu$.

\color{black}

\subsection{All-Island Irish Transmission System}
\label{case:aiits}

This section focuses on the computational efficiency of \eqref{eq:singlevector:sys}.
To this end, we employ a larger system, namely a 1,502-bus dynamic model of the \acf{aiits}.   The model has in total $1,629$ states and $9,897$ algebraic variables. Such a size makes the use of dense QZ-based eigensolvers to handle the formulation in \eqref{eq:singps} impractical, with the resulting pencil having dimensions 
$11,526\times11,526$.  
In contrast, the state matrix associated with the dense formulation \eqref{eq:ode:lin} is considerably smaller, at $1,629\times1,629$,
and therefore provides a useful reference for benchmarking.
Simulations in this section are executed using a computer with a 13th Gen Intel Core i7-13700, 2.10~GHz processor, 16~GB of RAM, and running a 64-bit Linux OS.

\begin{figure}[ht!]
  \centering
  \includegraphics[width=0.8\linewidth]{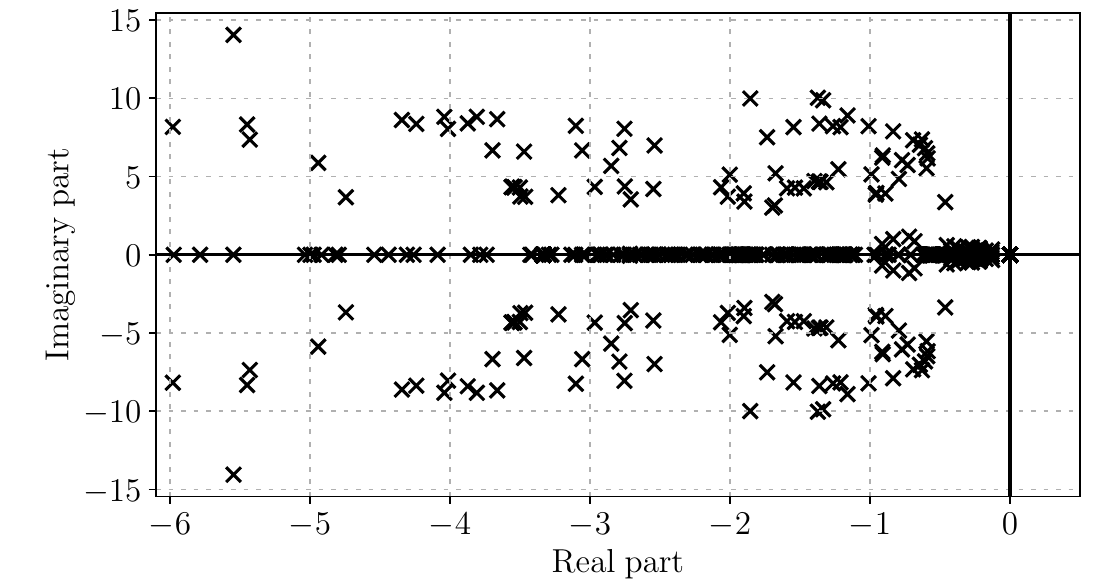}
  \caption{AIITS root locus.}
  \label{fig:battlefield}
\end{figure}

\begin{figure}[ht!]
  \centering
\includegraphics[width=0.8\linewidth]{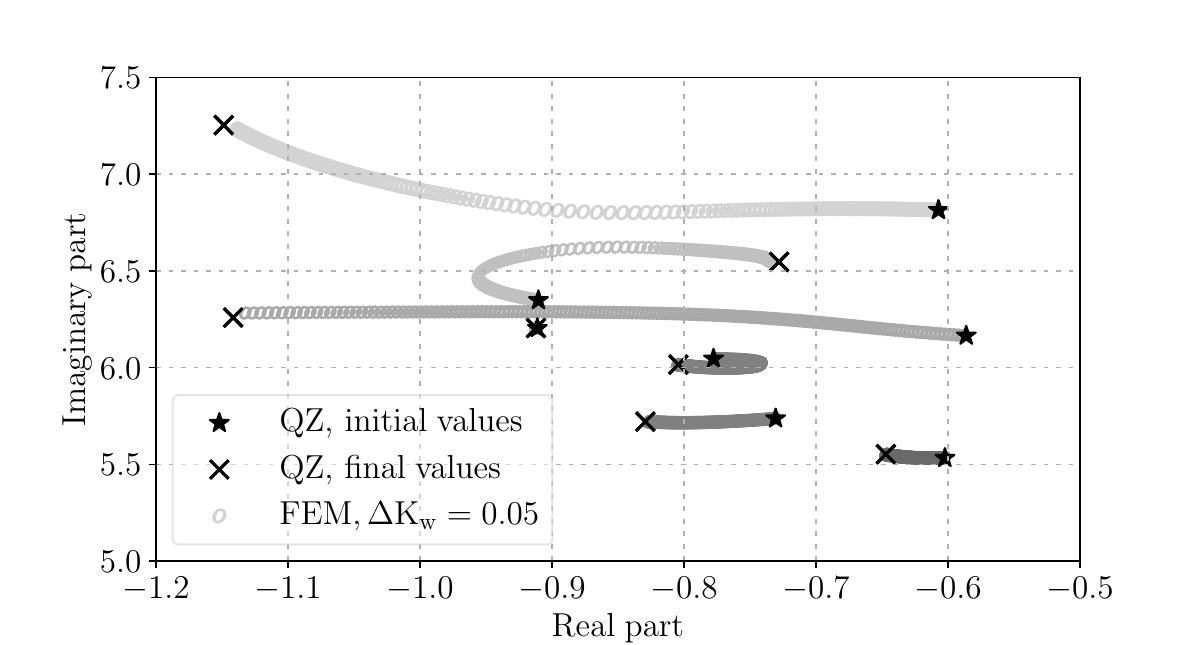}
  \caption{AIITS: Poorly damped electromechanical modes when the gains of the \acp{pss} are varied.}
  \label{fig:eir_GEP_PSS_Kw}
\end{figure}

As shown in the root locus of Fig.~\ref{fig:battlefield}, the high density of eigenvalues in the complex plane highlights that tracking a specific mode of interest can become a challenging task.  Firstly, we consider the formulation in \eqref{eq:singps} -- 
which results in an $11,526\times11,526$ system problem -- and focus on the region of the complex plane where poorly damped electromechanical modes are located.  We gradually increase the gains $\rm K_w$ of the $6$ \acp{pss} of the \ac{aiits} model from $1.5$ to $9.5$, and we sequentially trace a group of modes, shown in Fig.~\ref{fig:eir_GEP_PSS_Kw}. 
The results confirm that \eqref{eq:singlevector:sys} captures the eigenvalue trajectories with high precision.

\begin{figure}[ht!]
  \centering
\includegraphics[width=0.8\linewidth]{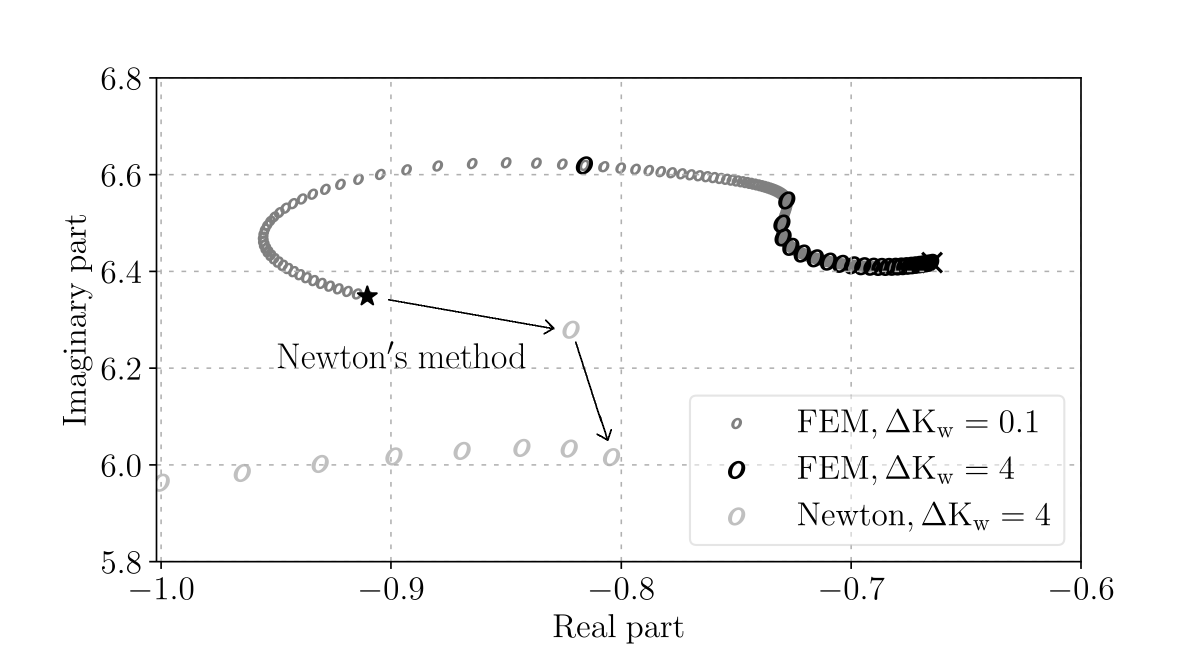}
  \caption{AIITS: Tracking an electromechanical mode: Newton vs integration-based continuation.}
  \label{fig:eir_GEP_PSS_Kw_NR}
\end{figure}

Unlike approaches that rely solely on Newton's method to compute eigenvalues at each parameter step (e.g.,~\cite{li2017eigenvalue}), the approach adopted in this paper is based on continuous integration and can, if needed, be combined with a Newton corrector to further improve accuracy.  This reduces the risk of divergence or convergence to incorrect values, particularly when large steps need to be used to boost computational efficiency.
As shown in Fig.~\ref{fig:eir_GEP_PSS_Kw_NR}, using a large step to track one of the system's electromechanical modes over the interval $\rm{K_w}=$[$1.5$, $41.5$] causes Newton's method, when applied directly to the eigenvalue problem, to deviate from the correct trajectory.
In contrast,~\eqref{eq:singlevector:sys} combined with a Newton corrector step consistently follows the correct eigenvalue trajectory, benefiting from more accurate intermediate estimates and improved convergence.

\begin{figure}[ht!]
  \centering
  \includegraphics[width=0.8\linewidth]{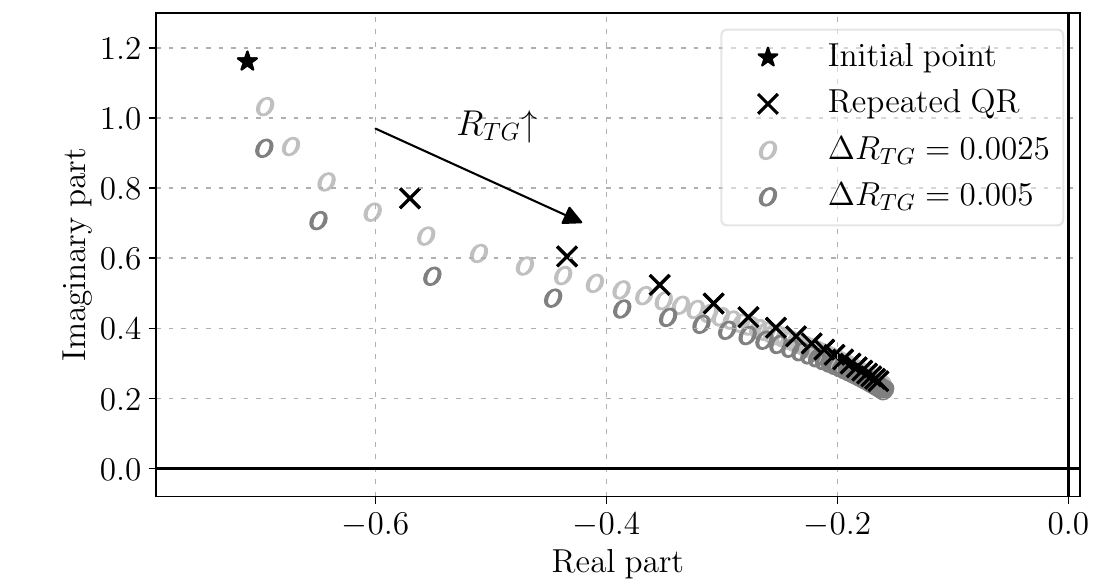}
  \caption{AIITS: Tracking of an electromechanical mode when the droop constant $R_{TG}$ of all \acp{sm} is varied.}
  \label{fig:eir_R_TG}
\end{figure}

\begin{figure}[ht!]
  \centering
  \includegraphics[width=0.8\linewidth]{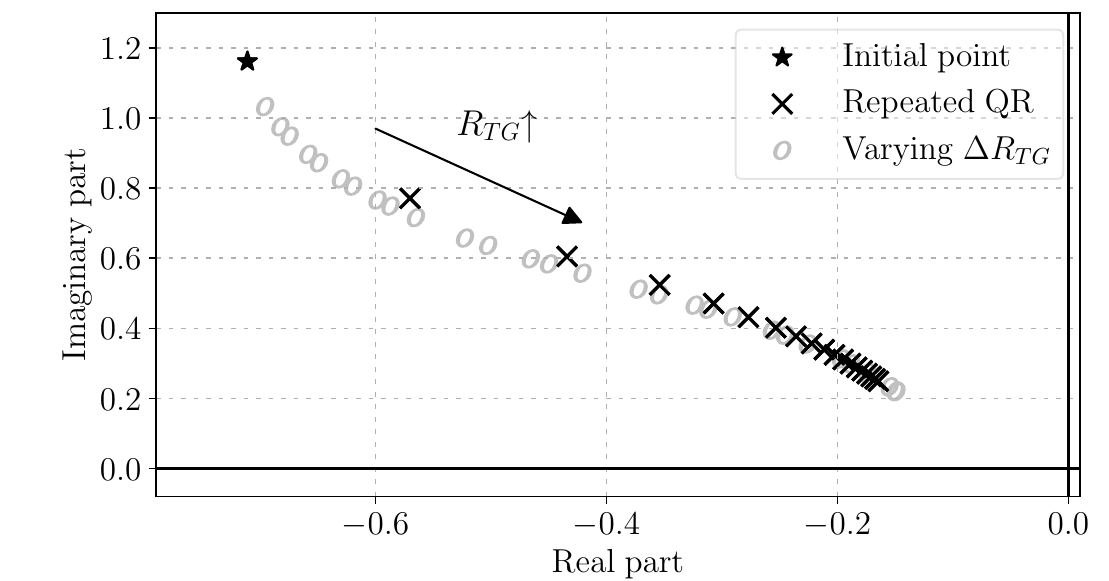}
  \caption{AIITS: Poorly damped mode tracking as $R_{TG}$ is varied.}
  \label{fig:eir_R_TG_var_h}
\end{figure}

Subsequently, we employ the formulation in \eqref{eq:ode:lin}. To further demonstrate the advantages of using an adaptive time step, we don't make use of a corrector step in figures \ref{fig:eir_R_TG} and \ref{fig:eir_R_TG_var_h}. \color{black}
A poorly damped mode of the system (initially located at $-0.71087 \pm \jj 1.16143$) is tracked as the \ac{tg} droop constants of all \acp{sm} are gradually increased from $0.02$ to $0.2$. The results for two different values of $\Delta R_{TG}$ are shown in Fig.~\ref{fig:eir_R_TG}.  As summarized in Table~\ref{tab:eir_R_TG}, 
\eqref{eq:singlevector:sys} is able to track the mode with high accuracy.
In terms of computational efficiency, the benchmark considered is the total time required to trace an eigenvalue over the range $[0.02, 0.2]$ using a step size of $\Delta R_{TG}=0.0025$. As shown in
Table~\ref{tab:aiits:fixed}, \eqref{eq:singlevector:sys} 
achieves a tenfold reduction of computation time compared to repeated QR factorizations.
The average time recorded for solving the power flow problem is $0.05$~s, with an additional $0.016$~s spent on post-processing tasks, including the construction of system matrices.
Therefore, the steady-state analysis step incurs an average overhead of $0.066$~s per iteration.

\begin{table}[!ht]
\caption{AIITS: Tracking accuracy for $R_{TG}$ variation.}
\label{tab:eir_R_TG}
\centering
\begin{tabular}{c|cc|cc}
\toprule
\centering
{}
 & \multicolumn{2}{c}{\textbf{$\Delta R_{TG} = 0.0025$}} & \multicolumn{2}{c}{\textbf{$\Delta R_{TG} = 0.005$}}\\
 \textbf{$R_{TG}$} & {Relative error} & {$\Delta \zeta \%$} & {Relative error} & {$\Delta \zeta \%$}\\
\midrule
$0.02$ & $0$ & $0$ & $0$ & $0$ \\
%\midrule
$0.05$ & $0.07146$ & $3.07406$ & $0.14665$ & $6.65217$\\
%\midrule
$0.075$ & $0.06434$ & $1.80305$ & $0.12820$ & $4.22215$\\
%\midrule
$0.1$ & $0.05955$ & $1.58114$ & $0.11782$ & $3.67346$\\
%\midrule
$0.15$ & $0.05233$ & $1.38234$ & $0.10325$ & $3.22334$\\
$0.2$ & $0.04665$ & $1.29035$ & $0.09214$ & $2.99423$\\
\bottomrule
\end{tabular}
\end{table}

\begin{table}[!ht]
\caption{AIITS: Tracking computational cost comparison.
}
\label{tab:aiits:fixed}
\centering
\begin{tabular}{c|c c c}
\toprule
\centering
{} & {Total time} & {After initial QR} & 
{1-step} 
\\
\midrule
{Repeated QR} & $459.54$~s & $453.67$~s & $6.30$~s \\
\midrule
{Eq.~\eqref{eq:singlevector:sys}} & $47.58$~s & $41.39$~s & $0.57$~s \\
\bottomrule
\end{tabular}
\end{table}

To further improve efficiency, we implement a simple adaptive step size strategy, where the integration step $\Delta R_{TG}$ is doubled or halved based on the Euclidean distance $|\Delta s|$ between two successive values of the tracked eigenvalue.  Specifically, the step is doubled if $|\Delta s| < 0.04$ and halved if $|\Delta s| > 0.08$. Figure~\ref{fig:eir_R_TG_var_h} shows the result of applying this adaptive strategy, starting from an initial step size of $0.0025$. The total computation time is significantly reduced from $47.58$~s (constant step) to $18.28$~s. As seen by comparing Figs.~\ref{fig:eir_R_TG} and \ref{fig:eir_R_TG_var_h}, this efficiency gain comes at a negligible cost in accuracy.

\section{Conclusion}
\label{sec:conclusions}

This paper focuses on continuation-based eigenvalue tracking of large-scale power system models.  A general formulation is presented that supports both dense and sparse matrix representations and enables efficient integration-based tracking of targeted eigenvalues.  Key implementation aspects are discussed in detail. The approach is first evaluated through a case study targeting the frequency regulation mode under various parameter variations. Computational efficiency and scalability are then discussed based on a 1,502-bus model of the Irish transmission system. Future work will explore extensions to more general classes of models, such as systems affected by measurement and communication delays. 

\section{Funding}

This work is supported by Research Ireland (SFI), under grant number 21/SPP/3756 through the NexSys Strategic Partnership Programme.

% \appendix
% \section[Derivation]{Derivation of \eqref{eq:phi:norm_dif}}
% \label{sec:appendix}

% The derivation of \eqref{eq:phi:norm_dif} 
% goes as follows.  Differentiation of \eqref{eq:phi:norm} with respect to $p \in \mathbb{R}$ gives:
% %
% \begin{equation}
% \label{eq:norm_deriv}
% \dot{\bfg{\phi}}\T\bfb E \bfg{\phi} + \bfg{\phi}\T\bfb E \dot{\bfg{\phi}} = -\bfg{\phi}\T \dot{\bfb E} \bfg{\phi}.
% \end{equation}
% %
% Let us denote as $\phi_i$ the $i$-th element of $\bfg{\phi}$ and as $e_{ij}$ the element of the $i$-th row and $j$-th column of $\bfb{E}$. 
% \color{red}
% By rewriting $\dot{\bfg{\phi}}\T\bfb E \bfg{\phi}$ as a sum, it is straightforward that:
% %
% \begin{equation}\label{eq:proof}
% \begin{aligned}
% \dot{\bfg{\phi}}\T\bfb E \bfg{\phi} &= \sum_{k=1}^r\sum_{l=1}^r \dot{\phi_l} e_{kl} \phi_l \\ 
% &=
% \sum_{k=1}^r \sum_{l=1}^r \phi_l e_{kl} \dot{\phi_l} 
% = 
% \bfg{\phi}\T \bfb E \dot{\bfg{\phi}}.
% \end{aligned}
% \end{equation}
% \color{black}
% %
% Combining \eqref{eq:proof} and leads to \eqref{eq:phi:norm_dif}. 

\bibliographystyle{IEEEtran}
\bibliography{refs}

\begin{thebibliography}{10}
\providecommand{\url}[1]{#1}
\csname url@samestyle\endcsname
\providecommand{\newblock}{\relax}
\providecommand{\bibinfo}[2]{#2}
\providecommand{\BIBentrySTDinterwordspacing}{\spaceskip=0pt\relax}
\providecommand{\BIBentryALTinterwordstretchfactor}{4}
\providecommand{\BIBentryALTinterwordspacing}{\spaceskip=\fontdimen2\font plus
\BIBentryALTinterwordstretchfactor\fontdimen3\font minus
  \fontdimen4\font\relax}
\providecommand{\BIBforeignlanguage}[2]{{%
\expandafter\ifx\csname l@#1\endcsname\relax
\typeout{** WARNING: IEEEtran.bst: No hyphenation pattern has been}%
\typeout{** loaded for the language `#1'. Using the pattern for}%
\typeout{** the default language instead.}%
\else
\language=\csname l@#1\endcsname
\fi
#2}}
\providecommand{\BIBdecl}{\relax}
\BIBdecl

\bibitem{milano2020eigenvalue}
F.~Milano, I.~Dassios, M.~Liu, and G.~Tzounas, \emph{Eigenvalue Problems in
  Power Systems}.\hskip 1em plus 0.5em minus 0.4em\relax CRC Press, 2020.

\bibitem{tajoli2023mode}
C.~Tajoli, G.~Tzounas, and G.~Hug, ``Mode-shape deformation of power system
  {DAEs} by time-domain integration methods,'' in \emph{2023 IEEE Belgrade
  PowerTech}.\hskip 1em plus 0.5em minus 0.4em\relax IEEE, 2023, pp. 1--6.

\bibitem{MAC_Pastor_2012}
M.~Pastor, M.~Binda, and T.~Harčarik, ``Modal assurance criterion,''
  \emph{Procedia Engineering}, vol.~48, pp. 543--548, 2012.

\bibitem{tzounas2020comparison}
G.~Tzounas, I.~Dassios, M.~Liu, and F.~Milano, ``Comparison of numerical
  methods and open-source libraries for eigenvalue analysis of large-scale
  power systems,'' \emph{Applied Sciences}, vol.~10, no.~21, p. 7592, 2020.

\bibitem{lehoucq1996deflation}
R.~B. Lehoucq and D.~C. Sorensen, ``Deflation techniques for an implicitly
  restarted {A}rnoldi iteration,'' \emph{SIAM Journal on Matrix Analysis and
  Applications}, vol.~17, no.~4, pp. 789--821, 1996.

\bibitem{stewart2002krylov}
G.~W. Stewart, ``A {K}rylov--{S}chur algorithm for large eigenproblems,''
  \emph{SIAM Journal on Matrix Analysis and Applications}, vol.~23, no.~3, pp.
  601--614, 2002.

\bibitem{sakurai2003projection}
T.~Sakurai and H.~Sugiura, ``A projection method for generalized eigenvalue
  problems using numerical integration,'' \emph{Journal of computational and
  applied mathematics}, vol. 159, no.~1, pp. 119--128, 2003.

\bibitem{ajjarapu1992continuation}
V.~Ajjarapu and C.~Christy, ``The continuation power flow: a tool for steady
  state voltage stability analysis,'' \emph{IEEE Transactions on Power
  Systems}, vol.~7, no.~1, pp. 416--423, 1992.

\bibitem{canizares1993point}
C.~A. Canizares and F.~L. Alvarado, ``Point of collapse and continuation
  methods for large {AC/DC} systems,'' \emph{IEEE Transactions on Power
  Systems}, vol.~8, no.~1, pp. 1--8, 1993.

\bibitem{Wen_2006}
X.~Wen and V.~Ajjarapu, ``Application of a novel eigenvalue trajectory tracing
  method to identify both oscillatory stability margin and damping margin,''
  \emph{IEEE Transactions on Power Systems}, vol.~21, no.~2, pp. 817--824,
  2006.

\bibitem{li2017eigenvalue}
C.~Li, G.~Li, C.~Wang, and Z.~Du, ``Eigenvalue sensitivity and eigenvalue
  tracing of power systems with inclusion of time delays,'' \emph{IEEE
  Transactions on Power Systems}, vol.~33, no.~4, pp. 3711--3719, 2017.

\bibitem{Cheng_2011}
C.~Luo and V.~Ajjarapu, ``Sensitivity-based efficient identification of
  oscillatory stability margin and damping margin using continuation of
  invariant subspaces,'' \emph{IEEE Transactions on Power Systems}, vol.~26,
  no.~3, pp. 1484--1492, 2011.

\bibitem{tdps1}
Q.~Mou, Y.~Xu, H.~Ye, and Y.~Liu, ``An efficient eigenvalue tracking method for
  time-delayed power systems based on continuation of invariant subspaces,''
  \emph{IEEE Transactions on Power Systems}, vol.~36, no.~4, pp. 3176--3188,
  2021.

\bibitem{zeng2016sensitivity}
Z.~Zeng, ``Sensitivity and computation of a defective eigenvalue,'' \emph{SIAM
  Journal on Matrix Analysis and Applications}, vol.~37, no.~2, pp. 798--817,
  2016.

\bibitem{dobson2001strong}
I.~Dobson, J.~Zhang, S.~Greene, H.~Engdahl, and P.~W. Sauer, ``Is strong modal
  resonance a precursor to power system oscillations?'' \emph{IEEE Transactions
  on Circuits and Systems I: Fundamental Theory and Applications}, vol.~48,
  no.~3, pp. 340--349, 2001.

\bibitem{wang2018analysis}
D.~Wang, L.~Liang, L.~Shi, J.~Hu, and Y.~Hou, ``Analysis of modal resonance
  between pll and dc-link voltage control in weak-grid tied vscs,'' \emph{IEEE
  Transactions on Power Systems}, vol.~34, no.~2, pp. 1127--1138, 2018.

\bibitem{semi:2016}
F.~{Milano}, ``Semi-implicit formulation of differential-algebraic equations
  for transient stability analysis,'' \emph{IEEE Transactions on Power
  Systems}, vol.~31, no.~6, pp. 4534--4543, Nov. 2016.

\bibitem{milano2022power}
F.~Milano, M.~Liu, M.~A. Murad, G.~M. J{\'o}nsd{\'o}ttir, G.~Tzounas, M.~Adeen,
  {\'A}.~Ortega, and I.~Dassios, ``Power system modelling as stochastic
  functional hybrid differential-algebraic equations,'' \emph{IET Smart Grid},
  vol.~5, no.~5, pp. 309--331, 2022.

\bibitem{kundur:94}
P.~Kundur, \emph{{Power System Stability and Control}}.\hskip 1em plus 0.5em
  minus 0.4em\relax New York: Mc-Grall Hill, 1994.

\bibitem{moebius}
I.~Dassios, G.~Tzounas, and F.~Milano, ``{ The M{\"o}bius transform effect in
  singular systems of differential equations},'' \emph{Applied Mathematics and
  Computation}, vol. 361, pp. 338--353, 2019.

\bibitem{robust}
------, ``Robust stability criterion for perturbed singular systems of
  linearized differential equations,'' \emph{Journal of Computational and
  Applied Mathematics}, vol. 381, p. 113032, 2021.

\bibitem{BOUHAMIDI2008687}
A.~Bouhamidi and K.~Jbilou, ``A note on the numerical approximate solutions for
  generalized {S}ylvester matrix equations with applications,'' \emph{Applied
  Mathematics and Computation}, vol. 206, no.~2, pp. 687--694, 2008.

\bibitem{ZHOU2008200}
B.~Zhou and G.-R. Duan, ``On the generalized {S}ylvester mapping and matrix
  equations,'' \emph{Systems \& Control Letters}, vol.~57, no.~3, pp. 200--208,
  2008.

\bibitem{JIN2020}
L.~Jin, J.~Yan, X.~Du, X.~Xiao, and D.~Fu, ``{RNN} for solving time-variant
  generalized {S}ylvester equation with applications to robots and acoustic
  source localization,'' \emph{IEEE Transactions on Industrial Informatics},
  vol.~16, no.~10, pp. 6359--6369, 2020.

\bibitem{SIMONCINI2016}
V.~Simoncini, ``Computational methods for linear matrix equations,'' \emph{SIAM
  Review}, vol.~58, no.~3, pp. 377--441, 2016.

\bibitem{web:39bus}
{Illinois Center for a Smarter Electric Grid (ICSEG)}, ``{IEEE 39-Bus
  System},''
  \href{http://publish.illinois.edu/smartergrid/ieee-39-bus-system/}{publish.illinois.edu/smartergrid/ieee-39-bus-system/}.

\bibitem{dome}
F.~Milano, ``{A Python-based software tool for power system analysis},'' in
  \emph{Proceedings of the IEEE PES General Meeting}, Jul. 2013.

\bibitem{Bernal_2016}
F.~Wilches-Bernal, J.~H. Chow, and J.~J. Sanchez-Gasca, ``A fundamental study
  of applying wind turbines for power system frequency control,'' \emph{IEEE
  Transactions on Power Systems}, vol.~31, no.~2, pp. 1496--1505, 2016.

\bibitem{Moeini_2016}
A.~Moeini and I.~Kamwa, ``Analytical concepts for reactive power based primary
  frequency control in power systems,'' \emph{IEEE Transactions on Power
  Systems}, vol.~31, no.~6, pp. 4217--4230, 2016.

\bibitem{2023enhancing}
F.~Milano, B.~Alhanjari, and G.~Tzounas, ``Enhancing frequency control through
  rate of change of voltage feedback,'' \emph{IEEE Transactions on Power
  Systems}, vol.~39, no.~1, pp. 2385--2388, 2023.

\bibitem{sun1990multiple}
J.~Sun, ``Multiple eigenvalue sensitivity analysis,'' \emph{Linear algebra and
  its applications}, vol. 137, pp. 183--211, 1990.

\bibitem{ORTEGA201837}
Álvaro Ortega and F.~Milano, ``Frequency control of distributed energy
  resources in distribution networks,'' \emph{IFAC-PapersOnLine}, vol.~51,
  no.~28, pp. 37--42, 2018.

\bibitem{ajjarapu2002continuation}
V.~Ajjarapu and C.~Christy, ``The continuation power flow: a tool for steady
  state voltage stability analysis,'' \emph{IEEE transactions on Power
  Systems}, vol.~7, no.~1, pp. 416--423, 2002.

\bibitem{milano2010continuation}
F.~Milano, ``Continuation power flow analysis,'' in \emph{Power system
  modelling and scripting}.\hskip 1em plus 0.5em minus 0.4em\relax Springer,
  2010, pp. 103--130.

\end{thebibliography}
\footnotesize

\end{document}